\documentclass[aps,prb,showpacs,twocolumn,superscriptaddress]{revtex4}
\usepackage{amsmath}
\usepackage{graphicx}
\usepackage{amsfonts}
\usepackage{amsbsy,amssymb,amsmath,bm}
\usepackage{graphicx}
\input{epsf}
\begin{document}
\title{Anisotropic superfluidity of two-dimensional excitons in a periodic potential}
\author{Yu. E. Lozovik}
\affiliation{Institute of Spectroscopy RAS, Moscow, Troitsk, Russia}
\affiliation{MIEM, National Research University Ц Higher School of Economics, Moscow }
\author{I. L. Kurbakov}
\affiliation{Institute of Spectroscopy RAS, Moscow, Troitsk, Russia}
\author{Pavel A. Volkov}
\affiliation{Theoretische Physik III, Ruhr-Universit\"at Bochum, D-44780 Bochum, Germany}

\date{\today}

\begin{abstract}
We study anisotropies of helicity modulus, excitation spectrum, sound velocity and angle-resolved luminescence spectrum in a two-dimensional system of interacting excitons in a periodic potential. Analytical expressions for anisotropic corrections to the quantities characterizing superfluidity are obtained. We consider particularly the case of dipolar excitons in quantum wells. For GaAs/AlGaAs heterostructures as well as MoS$_2$/hBN/MoS$_2$ and MoSe$_2$/hBN/WSe$_2$ transition metal dichalcogenide bilayers estimates of the magnitude of the predicted effects are given. We also present a method to control superfluid motion and to determine the helicity modulus in generic dipolar systems.
\end{abstract}

\maketitle

\section{\label{Intr} Introduction }
Bose systems below quantum degeneracy temperature are extensively studied now both theoretically and experimentally. In the last decades many phenomena theoretically predicted for weakly interacting Bose gases (Bose-Einstein condensation (BEC), Berezinskii-Kosterlitz-Thouless (BKT) transition \cite{BKT} to superfluid state in two-dimensional systems and related effects) have been confirmed by experiments with ultracold atoms or molecules in optical or magnetic traps \cite{atombec,prl091250401,nature441001118,pitstrbook}. Actively studied are also the superfluidity and BEC in systems of excitons\cite{Bbec} --- bound states of an electron and a hole in semiconductor nanostructures or exciton-polaritons --- mixed states of an exciton and a photon in an optical cavity \cite{nature443000409}. An important advantage of these systems over the
atomic/molecular ones is the low effective mass of the particles, which leads to degeneracy temperatures of the order of 1 K for excitons and tens of K for exciton-polaritons compared to nK for the atomic systems. However, their drawback is the finite lifetime of the particles which makes the inclusion of kinetic effects into consideration necessary.

Systems of indirect, dipolar excitons are of particular interest. Their important virtue is the considerably long lifetime due to small overlap of electron and hole wavefunctions which suppresses recombination. In contemporary studies the most widespread are 2D dipolar excitons \cite{Bbec,LY,2dde} in coupled quantum wells (QWs) \cite{LY} and wide single QWs \cite{K,F} in a polarizing normal electric field. After a pump pulse creating 2D dipolar excitons in a QW they thermalize locally quite fast\cite{prb042007434,prb059001625}. The excitons can then cool down to low temperatures\cite{prl086005608} due to the interaction with the semiconductor lattice\cite{prb053015834}, as their lifetime is sufficiently long \cite{K,nl0007001349}. Disorder (due to impurities and interface roughness) \cite{LB,rand}, inevitably present in semiconductors, is screened by the interexcitonic interactions\cite{screening} and is usually weak in wide single QWs\cite{K,prb046010193}. Fermionic (electron/hole) exchange effects in exciton-exciton interactions \cite{prb078045313} which destroy "boseness" $ $ of excitons \cite{prl099176403} and are disastrous \cite{pssc00302457,Tsuppr,prl110216407} for the BEC are suppressed at low densities if the dipole-dipole interaction of excitons is sufficiently strong
\cite{jetpl0660355,jetp10600296}. Excessive carriers \cite{trions} also suppressing BEC\cite{Tbec,prb075113301} can be compensated by injection of carriers with an opposite charge \cite{K}.

Finally, in the spatially separated\cite{prl096227402,prb083165308} continuous wave pumping regime cooled (charge compensated \cite{jetpl0960138}) excitons flow to the studied area of the QW. They can be cooled additionally by evaporative techniques \cite{prl103087403}. Due to the persistent inflow of cooled excitons even the degrees of freedom lowest in energy thermalize after a sufficiently long time\cite{prb056005306}. These considerations show that the BEC \cite{Tbec} (or at least mesoscopic BEC\cite{Tbec} or quasi-condensation\cite{prl097187402}) of 2D dipolar excitons is experimentally feasible \cite{Bbec,exexp}.

The remarkable experiments on exciton BEC have motivated rapid development of theoretical ideas \cite{extheor}. Many interesting effects have been predicted for exciton BECs: existence of non-dissipative electric currents \cite{jpc011l00483} and internal Josephson effect \cite{pla228000399}, roton instability \cite{prb090165430}, autolocalization \cite{Andreev} and (mesoscopic \cite{jetpl0790473}) supersolidity \cite{SSma,prl115075303}, BKT transition \cite{prl105070401} (crossover \cite{pla366000487}) and vortex formation \cite{prl100250401}, features of angle-resolved photoluminescence \cite{QClum} and nonlinear optical phenomena \cite{ssc126000269}, as well as topological excitons \cite{prb091161413}, dipole superconductivity \cite{sr0005011925}, and Casimir effect \cite{apl104162105}. Another interesting topic are exciton spin effects that have been predicted to lead to a multi-component BEC\cite{prl099176403}, which has been recently observed experimentally \cite{prl118127402}.

Important progress has been also achieved for other realizations: electron-electron bilayers in a quantizing magnetic field \cite{EE}, graphene bilayers \cite{graphen} (including realizations with a band gap\cite{bandgap}), topological insulator films \cite{topol} and cyclotron spin-flip excitons in wide GaAs single quantum wells with a 2D quantum-Hall electron gas at a filling factor of $\nu=2$. \cite{sr0005010354}

A high-temperature BEC phase of excitons \cite{14041418} can occur in 2D transition metal dichalcogenides \cite{TMD} based on \cite{nc0006006242,sci351000688} MoSe$_2$-WSe$_2$ or \cite{MoS2MoS2} MoS$_2$-MoS$_2$ bilayers if a hBN film is sandwiched in between two monolayers of a bilayer \cite{14041418}.

Various types of electrostatic \cite{jap099066104,apl097201106} and other \cite{prl096227402,other} traps for excitons are analogous to laser and magnetic traps for the atomic ensembles, e.g., flat traps \cite{flat} where during exciton lifetime an equilibrium density profile of excitons is formed \cite{prl103016403}. Specially designed electrostatic \cite{elst,apl100061103} and magnetic \cite{magn} lattices allow one to create an external periodic potential for excitons. Application of various voltage patterns \cite{prl106196806} can be used to create different types of potentials: confining \cite{jap099066104}, random one-dimensional and dividing the system with a barrier \cite{ol0032002466}. Superimposing two striped patterns \cite{apl085005830} or using more complex structures \cite{apl100061103} allows one to construct 2D potentials of various forms for excitons. Varying in-plane landscape of an electrode\cite{apl097201106} allows to create potentials\cite{jap117023108} and traps\cite{ol0040000589} of different types as well as potential energy gradients\cite{grad}.  There is also an another method of controlling excitons --- by creating deformation waves \cite{prl099047602}. Stationary periodic potential in this case can be realized with a standing acoustic wave.

BEC and superfluidity in periodic potentials is actively investigated for bosonic atoms. Considerable success in this field has been attained with atomic systems in laser traps \cite{pra084053601}: excitation spectrum measurements by Bragg spectroscopy were performed \cite{pra079043623}, a roton-maxon form of the excitation spectrum has been demonstrtated\cite{prl114055301}, processes similar to Bloch oscillations in crystals have been observed \cite{np0006000056}. From the theory side ground state and excitation spectrum have been calculated for models with a one-dimensional periodic potential\cite{epjd027000247} and effects similar to the ones for electrons in a crystal have been predicted, such as
Bloch oscillations \cite{pra058001480}. Formation of new phases by spontaneous symmetry breaking is also studied, in particular for dipole-type interacting systems \cite{SSBH,pra083023623}.

The majority of works, however, rely on the Bose-Hubbard model corresponding to a very strong periodic potential. This regime is hardly attainable for excitons for the following reasons. In the deep modulations regime, effective exciton mass is exponentially large \cite{meff}. Consequently, the superfluid transition temperature is exponentially low, and, even more importantly, the sensitivity to disorder \cite{prl069000644,pit-str} and free carriers increases greatly, setting very stringent constraints on the sample quality.

We consider thus the experimentally relevant case of excitons in a {\it weak} periodic potential which is opposed to the Bose-Hubbard approximation. For a macroscopically ordered excitonic system in the weak modulation regime we show that {\it anisotropic superfluidity} takes place, i.e. the dependence of the superfluid system observables on the direction in space. The superfluid density in this case turns out to be a tensor instead of a scalar. This precludes an interpretation of the superfluid density as being proportional to the number of particles in the condensate, which is always a scalar\cite{pr0104000576}.

One of the most famous examples of an anisotropic superfluid system is liquid $^3$He \cite{rmp047000331,volovik,nphys008000253}. Anisotropy there is caused by a condensate of atomic pairs with nonzero spin formed at low temperatures. Anisotropic superfluidity has been observed in atomic systems in optical lattices\cite{prl103165301,pra086043612} due to asymmetry $x\leftrightarrow y$ of the lattice potential. In this case anisotropy can be controlled : the reciprocal lattice vector sets the selected direction and the
amplitude of the potential governs the strength of anisotropic effects.

There are following observable effects related to anisotropic superfluidity:
\begin{itemize}
\item Changes in the vortex shape: non-dissipative currents around the vortex
core flow in ellipses rather than circles. In the 3D case this leads to an
elliptical form of vortex loops \cite{prl103165301}. Vortex cores become
elliptical as well \cite{prb084205113}.

\item Anisotropic optical coherence due to anisotropy of correlations
\cite{pra084033625}: visibility of the interference pattern in Young's
experiment \cite{Tbec} depends on the mutual orientation of the condensate
areas emitting light \cite{pra086053602}. In particular, coherence length may
depend on the direction in space \cite{prl105116402}.

\item In the strong anisotropy limit a finite system can effectively change
its dimensionality \cite{pra089023605}. Thus a long quasi-1D strip can behave
like a 2D system \cite{pra086043612} so that its transition to the normal
phase can be of BKT type. In 3D systems vortex loops can collapse to 2D
vortex-antivortex pairs, dissociation of which leads to a BKT-type transition
\cite{prl103165301,prb044004503}.

\item Finally, let us mention effects caused by the anisotropy of interparticle interactions rather than the superfluid density: anisotropies of sound\cite{pla376000480}, Landau critical velocity\cite{prl106065301}, dissipation\cite{njp005000050}, and of vortex properties: shape of the vortex core and intervortex interactions\cite{prl111170402}, as well as appearance of complex vortex structures\cite{sr0006019380}.
\end{itemize}

The aim of our work is to demonstrate the anisotropic superfluidity in a model of a weakly interacting two-dimensional Bose gas in a periodic potential, namely the effect of anisotropy of the potential on the superfluid motion characteristics and the elementary excitations. We will concentrate our attention on the dynamical/superfluid properties, and their anisotropic
character. In the present Article we consider a system of 2D dipolar excitons as an experimental realization of the model studied, though qualitative conclusions can be generalized to ultracold atomic systems. We present estimates of the magnitudes of the effects related to anisotropic superfluidity for the chosen physical realization.

The article is organized as follows. In Sec.\ref{anis} the tensor character of the superfluid density and anisotropic effects are discussed qualitatively. In Sec.\ref{main} a theoretical model is considered and analytical expressions for observable quantities are obtained. In Part\ref{phys} physical realizations of the studied model are described and qualitative manifestations of the anisotropic superfluidity are discussed. In Sec.\ref{estimation} we present estimates for the experimental effects
proposed. An outlook of the results obtained is presented in Sec.\ref{fin}.

\section{\label{anis} Anisotropic superfluidity }

In this work we consider three types of anisotropy: of sound velocity $C_s$ and of quantities related to linear response: superfluid mass density $\rho_s$ and helicity modulus ${\rm Y}_s$ \cite{pra008001111}. Anisotropy can have effect not on all the quantities, e.g., for bosons with an anisotropic interaction the critical velocity turns out to be anisotropic while the sound velocity is not\cite{prl106065301}.

Quantity $C_s$ can be deduced from the single-particle excitation spectrum $\varepsilon({\bf p})$ of the system. In the isotropic case energy of the excitations depends only on magnitude of the vector ${\bf p}$; we will show that in the presence of an anisotropic potential excitation spectrum as well as $C_s$ also depend on the direction of ${\bf p}$. A direct measurement of $C_s$ or $\varepsilon({\bf p})$ is needed to detect this type of anisotropy.

Quantities $\rho_s$ and ${\rm Y}_s$ are linear response coefficients connecting macroscopic flow parameters of the system such as current $\mathfrak{J}$, total momentum $\mathfrak{P}$ and energy $E$ to infinitesimal probe velocity ${\bf v}$ or momentum ${\bf P}$ transferred to each particle. In the isotropic case one has:

\begin{equation}
\begin{gathered}
\frac{\mathfrak{J}}{S} = \frac{\hbar}{2miS} \int
    \langle
        \hat \Psi^+ ({\bf r}) \nabla \hat \Psi ({\bf r}) - \left( \nabla \hat \Psi^+ ({\bf r}) \right) \hat \Psi ({\bf r})
    \rangle d {\bf r}
= {\rm Y}_s {\bf P},
\\
\frac{\mathfrak{P}}{S} = \frac{1}{S}\int
    \langle
        \hat \Psi^+ ({\bf r}) (-i \hbar \nabla) \hat \Psi ({\bf r})
    \rangle d {\bf r}
= \rho_s {\bf v},
\\
\frac{E}{S}=\frac{E({\bf P}=0)}{S}+\frac{{\rm Y}_s {\bf P}^2}{2} = \frac{E({\bf P}=0)}{S}+\frac{\rho_s {\bf v}^2}{2},
\end{gathered}
\label{j_p_defn}
\end{equation}
 where S is the system's area, ${\bf P} = m {\bf v}$ is the probe momentum, and a "phase twist" $ $\cite{pra008001111} condition is implied onto the field operator $\hat \Psi ({\bf r})$ : $\hat \Psi ({\bf r} + {\bf L}) = \exp{(iPL/\hbar)}\hat \Psi ({\bf r})$, where $L$ is the linear size of the system. It is natural to assume that definitions of $\rho_s$ and ${\rm Y}_s$ through momentum/current and energy coincide; though a general proof (including anisotropic case) for this statement has not been found by the authors but for the model we consider it follows from direct verification of the relation: $\mathfrak{J} = dE  /  d {\bf P}$.

Quantities $\rho_s$ and ${\rm Y}_s$ can be presented in the following form: $\rho_s = n_s m$ and ${\rm Y}_s = n_s / m$, where $m$ is the mass of particles and $n_s$ is the superfluid density. Temperature and {\it external fields} can lead to $n_s$ being smaller than full density $n$. This effect can be interpreted as being due to presence of a "normal component", which can be subject to dissipation and does not take part in the superfluid motion.

Let us discuss qualitatively what differences will be there for a system in an anisotropic external potential. For a single-particle problem it is known that a periodic potential leads to a change of the initial particle's mass to an effective mass tensor. Noticing that $\rho_s$ and ${\rm Y}_s$ play a role similar to mass in the expression for energy of the multi-particle system (coefficient with the square of velocity/momentum) we assume that $\rho_s$ and ${\rm Y}_s$ are also anisotropic tensor quantities. In this case for infinitesimal ${\bf P}$, ${\bf v}$ energy of the system per unit area will have the form

\begin{equation}
\begin{gathered}
\left. \frac{E}{S}\right|_{{\bf P}\rightarrow 0} \approx \frac{E({\bf P}=0)}{S}+\sum_{ij}\frac{(\rho_s)_{ij}v_i v_j}{2}
\\
=\frac{E({\bf P}=0)}{S}+\sum_{ij}\frac{({\rm Y}_s)_{ij}P_i P_j}{2},
\end{gathered}
\label{e_defn}
\end{equation}
or after substitution $P_1=|{\bf P}| \cos \phi; \; P_2=|{\bf P}| \sin \phi$ :
\begin{equation}
\begin{gathered}
\left. \frac{E}{S}\right|_{{\bf P}\rightarrow 0} \approx\frac{E({\bf P}=0)}{S}+\frac{{\rm Y}_s(\phi){\bf P}^2}{2}.
\end{gathered}
\label{e_defn_phi}
\end{equation}
Total momentum $\mathfrak{P}$ and current $\mathfrak{J}$ will be related to probe momentum and velocity in a similar way:
\begin{equation}
\begin{gathered}
\frac{\mathfrak{J}_i}{S} = \sum_j ({\rm Y}_s)_{ij} P_j,
\\
\frac{\mathfrak{P}_i}{S} = \sum_j (\rho_s)_{ij}v_j.
\end{gathered}
\label{j_dfn}
\end{equation}
In particular, it follows from (\ref{j_dfn}) that the probe velocity can be noncollinear with the total momentum. Quantity $n_s$ can be defined similarly to the isotropic case $\rho_{jl}=m^2 {\rm Y}^{jl}=m n_s^{jl}$; however, it will also be a tensor quantity which complicates the usual interpretation of the system as a mixture of "normal" $ $ and "superfluid" $ $ components. Let us mention that in the case when the initial mass of the particles is taken to be anisotropic (e.g., for a semiconductor with non-cubic lattice) $n_s$ can be a non-symmetric tensor unlike $\rho_s$ and ${\rm Y}_s$ for which symmetry follows from the definition (\ref{e_defn}).

Let us discuss methods to measure helicity modulus and superfluid mass density experimentally. To determine them one should transfer a uniform momentum ${\bf P}$ or velocity ${\bf v}$ to all the particles. Both options can be implemented: momentum can be transferred to excitons in crossed magnetic and electric fields by an adiabatic switch of the last one (see Sec.\ref{phys}) and velocity --- by setting the external potential into motion $V({\bf r}) \rightarrow V({\bf r - v} t)$. In the first case there will also be an effective addition to the exciton mass although it is negligibly small in sufficiently weak magnetic fields. Transforming the Hamiltonian to a moving reference frame one can show the equivalence of these approaches and that the transferred momentum is related to velocity through: ${\bf P}= m {\bf v}$, where $m$ is the exciton mass. What is left is to propose a method of measuring the system's current and momentum which is done in section \ref{phys} for a dipolar excitonic system.

Concluding the above one can see that to determine the parameters of anisotropic superfluidity one has to find mean values of Hamiltonian $\langle \hat H \rangle$ and current or momentum for the system in a state corresponding to a uniform motion with single-particle momentum ${\bf P}$ or velocity ${\bf v}$ and also the excitation spectrum $\varepsilon({\bf p})$.
\section{\label{main} Theoretical analysis}
Let us proceed to the theoretical model formulation. We will work in the following assumptions:
\begin{itemize}
\item Density of excitons is low enough and repulsion between the particles is sufficiently strong so that their composite fermionic structure can be ignored. Therefore, the excitons are considered to be strictly bosons.
\item Correlations are weak so that $(N-N_0)/N\ll1$, where $N$ is the total number of particles and $N_0$ is the number of particles in the condensate \cite{weakcorr}.
\item Modulation of the condensate profile by the periodic external field is weak. Specifically, we assume that $V_0$ is small compared to $\mu$.
\item Interparticle interactions do not involve spin and thus we ignore spin degrees of freedom for excitons \cite{ignoringspin}.
\end{itemize}

Thus, the considered system is a gas of weakly interacting bosons in an external periodic potential. Hamiltonian of the system (after change of variables from particle number to the chemical potential $\mu$) is:
\begin{equation}
\begin{gathered}
    \hat{H} - \mu \hat{N} =
    \int
        \hat{\Psi}^+({\bf r})
        \left(
            -\frac{\hbar^{2}}{2m}\Delta+V({\bf r})-\mu
        \right)
        \hat{\Psi}({\bf r})
    d {\bf r}
    +
\\
    \frac{1}{2}
    \int
        \hat{\Psi}^+({\bf r}) \hat{\Psi}^+({\bf r'} )
             U({\bf r}-{\bf r'})
        \hat{\Psi}({\bf r'})\hat{\Psi}({\bf r} )
    d{\bf r'}d {\bf r}
    ,
\end{gathered}
\label{ham}
\end{equation}
where $V({\bf r})=V_0\cos{\bf q}{\bf r}$ is the external potential and $U({\bf r}-{\bf r'})$ is the interparticle potential, which we consider to be symmetric under transformation ${\bf r}\leftrightarrow {\bf r'}$. We also assume the initial mass of the particles to be isotropic (it is true for, e.g., excitons in GaAs-based structures). In what follows we will consider the system at $T=0$.

For weakly correlated 2D bosons a standard Bogoliubov approach is applicable, provided one replaces bare interaction with an effective one arising from the summation of ladder diagrams \cite{loz-yud}. The condensate contribution to the energy of the system is:
\begin{equation}
\begin{gathered}
E_{cond} =
    \int
        \Phi^*({\bf r} )
        \left(
            -\frac{\hbar^{2}}{2m}\Delta+V({\bf r})
        \right)
        \Phi({\bf r} )
    d {\bf r}
    +
\\
    \frac{1}{2}
    \int
        \Phi^*({\bf r}) \Phi^*({\bf r'} )
             U({\bf r}-{\bf r'})
        \Phi({\bf r} )\Phi({\bf r'} )
    d{\bf r'} d {\bf r}.
\end{gathered}
\label{gp_en}
\end{equation}

Condensate wavefunction $\Phi({\bf r})$ satisfies the Gross-Pitaevskii equation (with $\mu$ being determined by normalization condition):
\begin{equation}
\begin{gathered}
    \left(
        -\frac{\hbar^{2}}{2m}\Delta+V({\bf r})-\mu+\int U({\bf r}-{\bf r'}) |\Phi({\bf r'})|^{2}d{\bf r'}
    \right)
    \Phi({\bf r})
    =0,
    \\
    \int |\Phi({\bf r})|^{2}d{\bf r} = N_0.
\end{gathered}
\label{gp}
\end{equation}
One should keep in mind that $N$ is the problem parameter while $N_0$ is not. They are connected through the relation $N= N_0 + N'$, where $N'$ is the number of particles depleted from the condensate which will be defined later in the article; both $N_0$ and $N'$ can depend on ${\bf P}$. We seek the solutions of equation (\ref{gp}) as a power series in $V_0$. For a uniformly moving condensate the zeroth order solution is taken as $const \cdot e^{i {\bf k} {\bf r}}$, where ${\bf k} = {\bf P} / \hbar$. Solution for $\Phi( {\bf r} ) $ and $\mu$ up to the second order in $V_0$ is:
\begin{equation}
\begin{gathered}
    \Phi({\bf r}) \approx \sqrt{n_0} e^{i {\bf k r} }\cdot
    \\
    \cdot \left( (1-V_0^2\Delta \Phi_0) -V_0\Phi_+ e^{i {\bf q r}}
        -V_0 \Phi_- e^{-i {\bf q r}} \right),
    \\
     \Phi_{\pm} =\frac{T \mp \alpha}{2 \kappa},\;
    \Delta \Phi_0 = \frac {\Phi_+^2+\Phi_-^2}{2},
    \\
   \mu = \frac {\hbar ^ 2 {\bf k} ^2}{2m} + n_0 U_{0} - V_0^2 T\frac{T^2 - \alpha^2}{2 \kappa^2},
\end{gathered}
\label{gpsol}
\end{equation}
where notations are introduced:
$n_0=N_0/S$,
$U_{{\bf q}} = \int U({\bf r}) e^{i {\bf q} {\bf r}} d {\bf r} $,
$U=n_0 U_{{\bf q}}$,
$T=\hbar^2 {\bf q}^2/2m$,
$\alpha=\hbar {\bf q} {\bf P}/m$,
$\kappa= T^2 +2TU - \alpha^2$.
We have omitted term $\sim e^{2 i {\bf q r}},\;e^{-2 i {\bf q r}}$ as their contribution to the quantities calculated further in text (condensate energy, excitation spectrum, etc.) is of higher, than second order in $V_0$. In the case ${\bf P} = 0$ solution takes the form:
\begin{equation}
\Phi({\bf r}) \approx \sqrt{n_0}  \left(1 -\frac{V_0^2}{4(T+2U)^2}- \frac{V_0}{T+2U}  \cdot \cos{\bf q} {\bf r} \right),
\label{gpsol1}
\end{equation}
which clearly demonstrates periodic modulations of condensate's density (with period determined by the external potential), i.e., diagonal long-range order.

\begin{figure}[h!]
\begin{center}
\includegraphics[width=8cm]{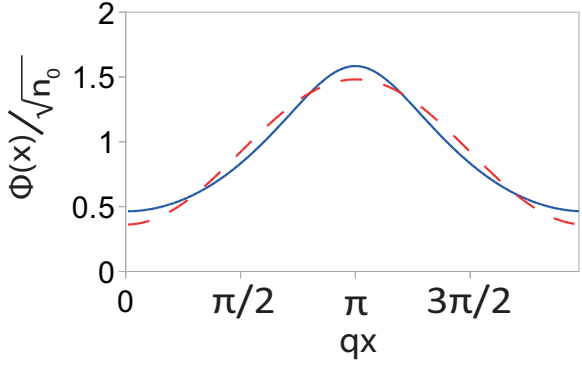}
\end{center}
\caption{Solution of the Gross-Pitaevskii equation (\ref{gp}) with the parameters taken for the MoS2/hBN/MoS2 structure (see Table \ref{t:struct}). Red dashed line is the approximate solution (\ref{gpsol1}), solid blue line is the numerical solution of (\ref{gp}).}
\label{fig:psi}
\end{figure}

In Fig.\ref{fig:psi} we compare the approximate solution (\ref{gpsol1}) with a full numerical solution. The approximation (\ref{gpsol1}) gives a reasonably good result (average relative error $<10\%$), particularly taking into account rather large anisotropic effects (see Table \ref{t:anissf}) for the same set of parameters. In what follows we use (\ref{gpsol}) to obtain closed analytical expressions for the quantities of interest.

Substituting the obtained solution into the expression for the energy (\ref{gp_en}) one has:
\begin{equation}
\begin{gathered}
\frac{E_{cond}}{S}=\frac{E_{cond}({\bf P}=0)}{S}+ \frac{ {\bf P}^2}{2m}n_0
\\
+ \frac{1}{2}(n_0^2-n_0^2({\bf P}=0))U_0
-\frac{m V_0^2 n_0 \alpha^2}{\hbar^2 {\bf q}^2 (T+2U)^2}.
\end{gathered}
\label{e_cond}
\end{equation}
It is also possible to calculate condensate contribution to the current $\mathfrak{J}$:
\begin{equation}
\begin{gathered}
\frac{\mathfrak{J}_{cond}}{S}=\frac{ {\bf P}}{m}n_0-\frac{V_0^2 n_0 \alpha \hbar {\bf q} }{m T (T+2U)^2}.
\end{gathered}
\label{j_cond}
\end{equation}

\begin{figure}[h!]
\begin{center}
\includegraphics[width=8cm]{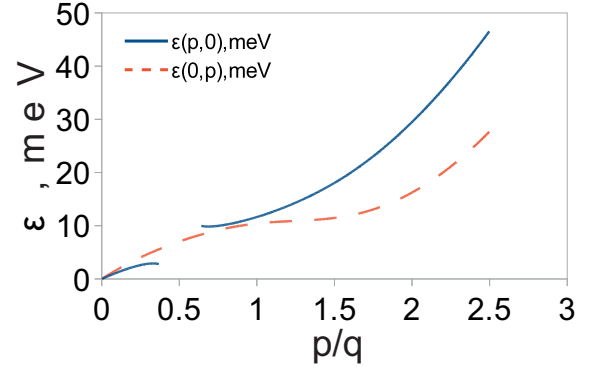}
\end{center}
\caption{Spectrum of elementary excitations (Eq.\ref{spectrum}) for the MoS2/hBN/MoS2 structure (see Table \ref{t:struct}). Shown are the excitation energies depending on the value of the momentum ${\bf p}$ measured in units of the reciprocal vector of the periodic potential ${\bf q}$ for ${\bf p}$ along and across ${\bf q}$. For ${\bf p}\parallel{\bf q}$ the region where the splitting effects become important (see text) is omitted.}
\label{fig:spectrum}
\end{figure}

Let us move on to the non-condensate part. We use the Bogoliubov transformation:
\begin{equation}
\begin{gathered}
       \hat{\Psi}'({\bf r} ) = \sum_l ( u_l({\bf r}) \hat{a}_l - v^*_l({\bf r}) \hat{a}^{+}_l ),
\end{gathered}
\label{bogol}
\end{equation}
where operators $\hat a_l$ and $\hat a^+_l$ satisfy Bose commutation relations. Diagonalizing the non-condensate Hamiltonian we obtain equations for $u_l({\bf r})$ and $v_l({\bf r})$ :
\begin{equation}
\begin{gathered}
\tilde T u_l({\bf r}) - \tilde U v_l({\bf r})=\varepsilon_l u_l({\bf r}),
\\
\tilde T v^*_l({\bf r}) - \tilde U u^*_l({\bf r})=-\varepsilon_l v^*_l({\bf r}),
\end{gathered}
\label{bogoleq}
\end{equation}
where

\begin{gather*}
\tilde T f ({\bf r})\equiv \left(-\frac{\hbar^{2}}{2m}\Delta+V({\bf r})-\mu \right.
\\
\left. + \int |\Phi({\bf r} ')|^2 U({\bf r} - {\bf r} ') d {\bf r}' \right) f({\bf r} )+
\\
    +\Phi({\bf r})\int \Phi^{*}({\bf r} ')U({\bf r} - {\bf r} ') f( {\bf r} ') d {\bf r}' ,
\\
\tilde U f ({\bf r}) \equiv
\Phi({\bf r})\int \Phi({\bf r}')U({\bf r} - {\bf r}') f( {\bf r} ') d {\bf r} '.
\end{gather*}
One can show that if the above equations are satisfied then the excitation Hamiltonian takes the form:
\[
\left(\hat H - \mu \hat N\right)' = \sum_l \varepsilon_l \hat a_l^+ \hat a_l - \sum_l \varepsilon_l \int |v_l ({\bf r})|^2 d {\bf r},
\]
and the mean non-condensate density is:
\[
n' =\frac{1}{S}\int \langle \hat \Psi'^{+}({\bf r})\hat \Psi'({\bf r}) \rangle d {\bf r}=
\frac{1}{S} \int \sum_l |v_l ({\bf r})|^2 d {\bf r}
\]
Equations (\ref{bogoleq}) are solved (and the excitation spectrum is found) approximately up to the second order in $V_0$. The Bogoliubov coefficients $u$, $v$ are then:
\begin{widetext}
\begin{equation}
\begin{gathered}
u_l({\bf r})= \frac{1}{\sqrt S} u_{{\bf p}}e^{i ({\bf k} + {\bf p}) {\bf r}}, \;
v_l({\bf r})= \frac{1}{\sqrt S}v_{{\bf p}} e^{i (-{\bf k} + {\bf p}) {\bf r}},
\\
u_{{\bf p}} =
u_{{\bf p}}^0  -
V_0 (u^+_{{\bf p}}e^{i {\bf q} {\bf r}}+u^-_{{\bf p}}e^{-i {\bf q} {\bf r}}) + V_0^2 \Delta u_0,
\\
v_{{\bf p}} =
v_{{\bf p}}^0-
V_0 (v^+_{{\bf p}}e^{i {\bf q} {\bf r}}+v^-_{{\bf p}}e^{-i {\bf q} {\bf r}}) + V_0^2 \Delta v_0,
\\
u_{{\bf p}}^0=\frac{1}{2}
\left(
    \sqrt{
            \frac{2m \varepsilon^0_{{\bf p}}}{\hbar^2{\bf p}^2}
        }+
    \sqrt{
        \frac{\hbar^2{\bf p}^2}{2m \varepsilon^0_{{\bf p}}}
        }
\right),
v_{{\bf p}}^0=\frac{1}{2}
\left(
    \sqrt{
            \frac{2m \varepsilon^0_{{\bf p}}}{\hbar^2{\bf p}^2}
        }-
    \sqrt{
        \frac{\hbar^2{\bf p}^2}{2m \varepsilon^0_{{\bf p}}}
        }
\right)  ,
\\
\varepsilon^0_{{\bf p}}=\sqrt{\left( \frac{\hbar^2 {\bf p}^2}{2m} \right)^2 +
n_0 U_{{\bf p}} \frac{\hbar^2 {\bf p}^2}{m}},
\end{gathered}
\label{bogolsol}
\end{equation}
where $v_{{\bf p}}^\pm$ and $u_{{\bf p}}^\pm$ are given by:
\begin{equation}
\begin{gathered}
u^\pm_{{\bf p}}=
\frac
{A_\pm(T_\pm+n_0 U_{{\bf p} \pm {\bf q}}+\varepsilon^0_{{\bf p}} \mp \alpha)+B_\pm n_0 U_{{\bf p} \pm {\bf q}}}
{(\varepsilon^{0}_{{\bf p} \pm {\bf q}})^2-(\varepsilon^0_{{\bf p} }\mp \alpha)^2},
\\
v^\pm_{{\bf p}}=
\frac
{A_\pm n_0 U_{{\bf p} \pm {\bf q}}+B_\pm(T_\pm+n_0 U_{{\bf p} \pm {\bf q}}-\varepsilon^0_{{\bf p}} \pm \alpha)}
{(\varepsilon^{0}_{{\bf p} \pm {\bf q}})^2-(\varepsilon^0_{{\bf p} }\mp \alpha)^2},
\\
A_\pm= u^0_{{\bf p}} \frac{T^2-\alpha^2}{2 \kappa} - \frac{T n_0 (U_{{\bf p}}+U_{{\bf p} \pm {\bf q}})}{2\kappa}f_-
\pm \alpha n_0 \frac{U_{{\bf p}}f_--U_{{\bf p} \pm {\bf q}}f_+}{2\kappa},
\\
B_\pm= v^0_{{\bf p}} \frac{T^2-\alpha^2}{2 \kappa} + \frac{T n_0 (U_{{\bf p}}+U_{{\bf p} \pm {\bf q}})}{2\kappa}f_-
\pm \alpha n_0 \frac{U_{{\bf p}}f_-+U_{{\bf p} \pm {\bf q}}f_+}{2\kappa},
\\
f_+=\sqrt{\frac{2m \varepsilon^0_{\bf p}}{\hbar^2 {\bf p}^2}} ,\;
f_-=\sqrt{\frac{\hbar^2{\bf p}^2 }{2m \varepsilon^0_{\bf p}}} ,
\end{gathered}
\label{uv}
\end{equation}
where $T_\pm=\hbar^2 ({\bf p} \pm  {\bf q})^2 /2m$.
The second order corrections $\Delta v^0_{\bf p}$ and $\Delta u^0_{\bf p}$ are:
\begin{equation}
\begin{gathered}
2 v^0_{\bf p} \Delta v^0_{\bf p} = \frac{n_0 U_{\bf p}}{2 (\varepsilon^0_{\bf p})^2}(A  v^0_{\bf p} + B u^0_{\bf p})
+C (v_{\bf p}^0)^2,
\\
2 u^0_{\bf p} \Delta u^0_{\bf p} = \frac{n_0 U_{\bf p}}{2 (\varepsilon^0_{\bf p})^2}(A  v^0_{\bf p} + B u^0_{\bf p})
+C (u_{\bf p}^0)^2,
\\
A= \frac{T^2-\alpha^2}{2 \kappa}(u^+_{{\bf p}}+u^-_{{\bf p}})-\frac{T^2-\alpha^2}{2 \kappa^2}T u_{{\bf p}}^0
+n_0 U_{{\bf p}} \left( (\Phi_+^2 + \Phi_-^2)f_- + \Phi_+v^-_{{\bf p}}+\Phi_-v^+_{{\bf p}}-\Phi_+u^+_{{\bf p}}-\Phi_-u^-_{{\bf p}} \right)+
\\
+n_0U_{{\bf p}+{\bf q}}\Phi_-(\Phi_+v_{{\bf p}}^0 - \Phi_-u_{{\bf p}}^0 + v^+_{{\bf p}} -  u^+_{{\bf p}})
+n_0 U_{{\bf p}-{\bf q}}\Phi_+(\Phi_-v_{{\bf p}}^0 - \Phi_+u_{{\bf p}}^0 + v^-_{{\bf p}} -  u^-_{{\bf p}}),
\\
B= \frac{T^2-\alpha^2}{2 \kappa}(v^+_{{\bf p}}+v^-_{{\bf p}})-\frac{T^2-\alpha^2}{2 \kappa^2}T v_{{\bf p}}^0
+n_0 U_{{\bf p}} \left( -(\Phi_+^2 + \Phi_-^2)f_- + \Phi_+u^+_{{\bf p}}+\Phi_-u^-_{{\bf p}}-\Phi_+v^-_{{\bf p}}-\Phi_-v^+_{{\bf p}} \right)+
\\
+n_0 U_{{\bf p}+{\bf q}}\Phi_+(\Phi_-u_{{\bf p}}^0 -\Phi_+v_{{\bf p}}^0 + u^+_{{\bf p}} -  v^+_{{\bf p}})
+n_0 U_{{\bf p}-{\bf q}}\Phi_-(\Phi_+u_{{\bf p}}^0 - \Phi_-v_{{\bf p}}^0 + u^-_{{\bf p}} -  v^-_{{\bf p}}),
\\
C=(v^+_{\bf p}+u^+_{\bf p})(v^+_{\bf p}-u^+_{\bf p})+(v^-_{\bf p}+u^-_{\bf p})(v^-_{\bf p}-u^-_{\bf p}).
\end{gathered}
\label{deltauv}
\end{equation}
\end{widetext}
The excitation spectrum is given by:
\begin{equation}
\begin{gathered}
\varepsilon_{{\bf p}}=\varepsilon^0_{{\bf p}} +\hbar {\bf p} \frac{\bf P}{m} +
V_0^2 \Delta \varepsilon_{{\bf p}},
\\
\Delta \varepsilon_{{\bf p}} = - ( u^0_{{\bf p}}A+v^0_{{\bf p}}B).
\end{gathered}
\label{spectrum}
\end{equation}
In Fig. \ref{fig:spectrum} we present a spectrum for one of the structures described in Sec.\ref{phys}-\ref{estimation}. At low {\bf p} it has the linear Bogoliubov form with anisotropic sound velocities (see also Table \ref{t:anissf}). For ${\bf p}\parallel{\bf q}$ one can see a characteristic flattening starting near $p \approx q/2$. This is a signature of the spectrum splitting near the edge of the Brillouin zone defined by the external potential. The expression (\ref{spectrum}) is not applicable in this region. In what follows we consider the superfluid properties of the system, which are determined by the low-${\bf p}$ part of the spectrum, so we do not consider the effects induced by the splitting and the corresponding region is omitted in the figure. Another interesting detail is a developing roton-minimum-like feature for ${\bf p}\perp{\bf q}$. However, it is clearly far from an instability and we do not study this feature in detail, as it does not affect the superfluid properties that we consider below.
For the depleted density we have:
\begin{equation}
\begin{gathered}
n'=n-n_0= \frac{1}{(2 \pi)^2} \int d {\bf p} \left\{ |v_{\bf p}^0|^2+ \right.
\\
\left. V_0^2|v_{\bf p}^+|^2+V_0^2|v_{\bf p}^-|^2+ 2 V_0^2 v_{\bf p}^0 \Delta v_{\bf p}^0 \right\} =
\\
\frac{1}{(2 \pi)^2} \int d {\bf p}  \left\{  |v_{\bf p}^0|^2+
V_0 ^2 n_0 U_{{\bf p}} \frac{A v_{\bf p}^0 + B  u_{\bf p}^0}{2(\varepsilon^{0}_{\bf p})^2} \right.
\\
+V_0^2(|v_{\bf p}^+|^2+|v_{\bf p}^-|^2) u^2_0 ({\bf p})
\\
\left. -V_0^2(|u_{\bf p}^+|^2+|u_{\bf p}^-|^2) v^2_0 ({\bf p}) \right\}  .
\end{gathered}
\label{nc_dens}
\end{equation}
Thus, taking into account $N_0+N'=N$ we have an equation for $N_0$.
For an arbitrary potential the integral in (\ref{nc_dens}) cannot be evaluated analytically; however, one can study its convergence. For convergence at ${\bf p}\to0$ and ${\bf p}\to\infty$ it is sufficient for
the potential $U_{\bf p}$ to be finite. In the second order in $V_0$ there are also two problematic points where $\varepsilon_{{\bf p}} =\varepsilon_{{\bf p} + {\bf q}} \mp \alpha$. For ${\bf P} = 0$ ($\alpha=0$) the singularity can be integrated in the principal value sense without taking splitting into account. In the case when ${\bf P} \neq 0$ this is not possible because the singularity is $1/x^2$. However, taking splitting into account will certainly lead to a finite result. Then it turns out that condensate depletion in the system is finite and consequently there is a non-zero condensate fraction for the weakly correlated system\cite{nc_finite}.

Contribution of the depleted particles to the energy of the system takes the form:
\begin{equation}
\begin{gathered}
\langle \hat H ' \rangle = \mu N'
- \sum_l \varepsilon_l \int |v_l ({\bf r})|^2 d {\bf r} =
\\
\mu n' S
-\frac{S}{(2 \pi)^2} \int d {\bf p}
\varepsilon({\bf p}) \left\{
|v_{\bf p}^0|^2+
V_0 ^2 n_0 U_{{\bf p}} \frac{A v_{\bf p}^0 + B  u_{\bf p}^0}{2(\varepsilon^{0}_{\bf p})^2}
\right.
\\
\left.
+V_0^2(|v_{\bf p}^+|^2+|v_{\bf p}^-|^2) (u_{\bf p}^0)^2
-V_0^2(|u_{\bf p}^+|^2+|u_{\bf p}^-|^2) (v_{\bf p}^0)^2
\right\}.
\end{gathered}
\label{e_nc}
\end{equation}
Let us discuss the convergence of expression (\ref{e_nc}). The first term diverges at large momenta for potentials finite at ${\bf p}\to\infty$; however, this is resolved by taking the second Born approximation for the interaction potential into account. For convergence of the remaining terms at ${\bf p}\to\infty$ it is necessary to put a more stringent condition than before on $U_{\bf p}$:
$2U_{\bf p}-U_{\bf p+q}-U_{\bf p-q}=O\left( \frac{1}{|{\bf p}|^{\beta}} \right),\; \beta>0,\; |{\bf p}|\rightarrow \infty$. However, even for a delta-function potential it is clearly fulfilled. For real physical potentials this condition is met because of the finite size of the particles, inside which large repulsion forces act, e.g., in a model potential for dipolar excitons constructed in accord with the results of a numerical simulation (see App. \ref{Ueff}).

It is also not possible to give an analytical answer for (\ref{e_nc}) in the general case but it is possible to draw conclusions on the dependence of $\langle \hat H ' \rangle$ on ${\bf P}$. It turns out\cite{alpha_dep} that for small ${\bf P}$:
\begin{equation}
\begin{gathered}
\langle \hat H ' \rangle \approx \langle \hat H ' \rangle \mid_{{\bf P}\rightarrow 0} + a_{E'} \alpha^2,
\\
n' \approx n'\mid_{{\bf P}\rightarrow 0} + a_{n'} \alpha^2,
\end{gathered}
\label{nc_decomp}
\end{equation}
where $a_{E'}$ and $a_{n'}$ are constants. Now one can show using the expression (\ref{e_cond}) that the total energy of the system takes the form:
\begin{equation}
\begin{gathered}
\left. \frac{E}{S} \right| _{{\bf P}\rightarrow 0} \approx
\\
\frac{E({\bf P}=0)}{S} + \frac{{\bf P}^2}{2 m}n + \frac{a_E}{2} ({\bf q} {\bf P})^2=
\\
\frac{E({\bf P}=0)}{S}+\sum_{ij} \left(\frac{n}{m} \delta_{ij}+ a_E q_i q_j \right)P_i P_j/2.
\end{gathered}
\label{energy}
\end{equation}
where $a_E$ is a constant. Comparing with helicity modulus definition (\ref{e_defn}) we have: $({\rm Y}_s)_{ij}=( n_0/m \delta_{ij}+ a_E q_i q_j)$. To obtain superfluid mass density $\rho_s$ one should express momentum ${\bf P}$ through velocity and substitute into (\ref{energy}). Comparing with the definition we have: $(\rho_s)_{ij}= m^2({\rm Y}_s)_{ij} = ( n_0 m \delta_{ij}+ a_E m^2 q_i q_j)$. Thus we have shown that the system under study is superfluid and the helicity modulus and the superfluid mass density are anisotropic tensor quantities.

One can come to similar conclusions starting from definitions (\ref{j_dfn}). The non-condensate contribution to the total current is then given by:
\begin{gather*}
\frac{\mathfrak{J}'}{S} = \frac{\hbar}{2miS} \int
    \langle
        \hat \Psi '^+  ({\bf r}) \nabla \hat \Psi' ({\bf r}) - \left( \nabla \hat \Psi '^+ ({\bf r}) \right) \hat \Psi' ({\bf r})
    \rangle d {\bf r}
    =
    \\
\frac{{\bf P}}{m}n'
+\frac{1}{m(2 \pi)^2} \int d {\bf p}
{\bf p} \left\{ |v_{\bf p}^0|^2+ V_0^2|v_{\bf p}^+|^2+V_0^2|v_{\bf p}^-|^2+
\right.
\\
\left.+ 2 V_0^2 v_{\bf p}^0 \Delta v_{\bf p}^0 \right\}
+\hbar {\bf q} V_0^2 \left\{ |v_{\bf p}^+|^2-|v_{\bf p}^-|^2 \right\}
\end{gather*}

From the calculations above we can quantitatively discuss anisotropy of sound velocity. It can be obtained from the spectrum (\ref{spectrum}) as $C_s = \left.\partial\varepsilon_{\bf p}/\partial p\right|_{p,P=0}$ yielding:
\begin{equation}
\begin{gathered}
C_s = \sqrt{\frac{n_0 U_0}{m}} \left(1-\frac{V_0^2}{(T+2U)^2}\cos^2 \varphi +\frac{V_0^2(T-U)  U_{\bf q}}{U_0 (T+2U)^3} \right),
\end{gathered}
\label{csnd}
\end{equation}
where $\varphi$ is the angle between ${\bf p}$ and ${\bf q}$. One can see the presence of an anisotropic contribution in (\ref{csnd}). It is interesting to note that corrections to (\ref{spectrum}) due to periodic potential become unimportant as $p\rightarrow \infty$. One can prove that $\Delta \varepsilon_{{\bf p}} \rightarrow const$ if $|{\bf p}| \rightarrow \infty$, becoming negligible compared to $\varepsilon_0({\bf p})$.

To demonstrate the physics of anisotropic superfluidity we would also like to calculate ${\rm Y}_s$. The calculation can be carried out properly with the help of a relation\cite{pit-str}:
\begin{equation}
\begin{gathered}
C_s(\phi)=\sqrt{\frac{d \mu}{dn}{\rm Y}_s(\phi)},
\\
\frac{d \mu}{dn}=\frac{d \mu}{dn_0} \left(\frac{dn}{dn_0}\right)^{-1},
\end{gathered}
\label{pit-str}
\end{equation}
where $\phi$ is the angle between ${\bf q}$ and ${\bf v}$ and all the quantities are taken at ${\bf P}=0$.

Condensate depletion is: $n' \approx n_0 m U_0/(2 \pi \hbar^2)$, i.e., we have $n_0(dn/dn_0)=n$. One obtains using (\ref{csnd}):
\begin{equation}
{\rm Y}_s(\phi)=\frac{n}{m}\left(1-2 \frac{V_0^2}{(T+2U)^2}\cos^2 \phi\right).
\label{y_sq}
\end{equation}
As is discussed in Sec.\ref{anis} (see Eqs.(\ref{e_defn}),(\ref{j_dfn})), the helicity modulus is in general case a tensor. To illustrate this we rewrite Eq.(\ref{y_sq}) in tensor form\cite{anis-estim}:
\begin{equation}\label{Ysmatr}
||{\rm Y}_s||=\frac nm\left(\begin{array}{cc}
1-2\Delta_s\;&0\\0\;&1
\end{array}\right),
\end{equation}
where $x$ axis is along the wavevector and the anisotropy parameter $\Delta_s$ is defined as follows:
\begin{equation}\label{Deltas}
\Delta_s=\frac{V_0^2}{(T+2U)^2}.
\end{equation}
It is useful to consider a particular case where the form of ${\rm Y}_s(\phi)$ is known. Let us consider a case when $|{\bf q}|\rightarrow \infty$, $U_{\bf p}=U_0=const$. In this case it should be possible to explain anisotropy of ${\rm Y}_s$ from an effective mass point of view. The effect of an external potential $V({\bf r})$ is then reduced to substitution of the initial mass with a tensor $m_{ij}$ determined from a single-particle problem in the potential $V({\bf r})$. With accuracy up to the second order in $V_0$ in coordinates where $x$ axis is along ${\bf q}$, $m_{ij}$ has the form:
\[
m_{ij}=m_i \delta_{ij}, \; m_1= m+ 8 m^3 V_0^2/(\hbar {\bf q})^4 , \; m_2=m.
\]

Energy calculation for motion with probe momentum ${\bf P}$ leads us to:
\begin{gather*}
\left. \frac{E}{S} \right| _{{\bf P}\rightarrow 0} \approx
\frac{E({\bf P}=0)}{S} + \frac{1}{2} \sum_{ij} (m)^{-1}_{ij} P_i P_j
\end{gather*}
Then we have from the definition (\ref{e_defn_phi}):
\begin{gather*}
{\rm Y}_s(\phi)=\frac{n}{ m+8 m^3 V_0^2/(\hbar {\bf q})^4 }\cos^2 \phi + \frac{n_s}{m} \sin^2 \phi \approx
\\
 \frac{n}{m}(1-8m^2V_0^2/(\hbar {\bf q})^4 \cos^2 \phi),
\end{gather*}
which coincides with (\ref{y_sq}).

Thereby we have shown presence of a BEC, superfluidity and a diagonal long-range order in the system and demonstrated anisotropy of superfluid properties. However, despite the occurence of BEC, superfluidity and diagonal long-range
order the system {\it it is not a real supersolid}. Indeed, the diagonal long-range order does not involve a possibility of static deformations because the order is created artificially by the external potential. In a true supersolid, on the contrary, the modulations emerge due to self-organization, caused by an instability of the homogeneous phase with respect to formation of periodic (crystalline) modulation in the density profile, static deformations being possible.

Results obtained can be generalized for spatial lattices additively for energy, spectrum and condensate depletion because all of the equations studied were linearized and for energy and current cross-terms stemming from different modulation wavevectors vanish after integration. For a limit $|{\bf q}|\rightarrow \infty$, $U_{\bf p}=U_0=const$ and a square or triangular lattice one can see that anisotropy of helicity modulus and superfluid mass density is absent. It is also convenient to generalize results for three-dimensional systems; the only peculiarity is an additional constraint $2U_{\bf p}-U_{\bf p+q}-U_{\bf p-q}=O\left( \frac{1}{|{\bf p}|^{1+\beta}} \right),\; \beta>0,\; |{\bf p}|\rightarrow \infty$ for the non-condensate energy to converge.

A straightforward generalization can be also obtained in the case when there is an intrinsic mass anisotropy. An answer for the helicity modulus can be obtained for this case by transforming the tensor (\ref{Ysmatr}) to the frame where the mass tensor is diagonal and a change in the definition of $\Delta_s$:
\begin{equation}
||{\rm Y}_s||=\left(\begin{array}{cc}
n_0/m_1-A_1\;&-A_{12}\\-A_{12}\;&n_0/m_2-A_2\end{array}\right),
\label{YsmatrAnisMass}
\end{equation}
where
$$
A_1=2 \frac{V_0^2\hbar^2q_1^2n_0}{T'(T'+2U)m_1^2},\;\;\;\;
A_2=\frac{V_0^2\hbar^2q_2^2n_0}{T'(T'+2U)m_2^2},
$$
$$
A_{12}=\frac{V_0^2\hbar^2q_1q_2n_0}{T'(T'+2U)m_1m_2},\;\;\;\;
T'=\frac{\hbar^2q_1^2}{2m_1}+\frac{\hbar^2q_2^2}{2m_2},
$$
with $q_1$ and $q_2$ being the components of ${\bf q}$ in the principal axes frame of the mass tensor and $m_1$, $m_2$ are its eigenvalues.

\section{\label{phys} Physical realization }
\begin{figure}[h!]
\begin{center}
\includegraphics[width=8cm]{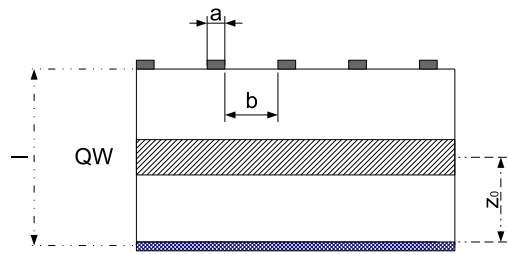}
\end{center}
\caption{Cross-sectional sample scheme. Dark areas on the top are the metallic stripes of the upper electrode, shaded area in the middle is the QW region, shaded area at the bottom is the lower electrode}
\label{fig:exp}
\end{figure}

To observe the effects described in Sec.\ref{main} we propose to use a system of dipolar excitons in a QW (or coupled QWs) in an external electrostatic field created by electrodes sputtered on the sample. A principal scheme of the realization discussed is shown in Fig.\ref{fig:exp}.
The bottom electrode is a flat layer of a doped semiconductor. The top one consists of periodically arranged (with period $a+b$)  metallic stripes of width $a$, with the separation between them being $b$. We assume the thickness of the stripes to be small enough for the top electrode to be semitransparent for recombination radiation of photons.

Inhomogeneous electrostatic field appearing when a voltage is applied to the electrodes creates a periodic potential for the excitons in the QW plane by interacting with their dipole moment. The period $\lambda$ of the potential depends on the overall period of the top electrode $a+b$ as well as on the distribution of voltages on them (in case it is not uniform). Magnitude of the applied voltage determines the amplitude of potential oscillations $V_0$ as well as the constant component ${\bf E}_{\rm av}=\{0,0,E_z^{\rm av}\}$ of the electric field. The latter determines the dipole moment of the excitons in single QWs and their lifetime in the radiation zone \cite{prb042009225}.

Proposed realization has two important limitations. First, for observing excitons in a superfluid state it is necessary for them to be in thermodynamic equilibrium. This happens only if exciton's relaxation time is not larger than their lifetime determined by recombination processes. The second limitation arises because of the presence of an electric field component parallel to the QW plane in electrostatic traps. In the case when dipole energy becomes on the order of exciton binding energy electron and hole may tunnel to an unbound state which leads to large leakage and prohibits observation of condensation \cite{prb072075428}.

\begin{figure}[h!]
\includegraphics[width=6cm]{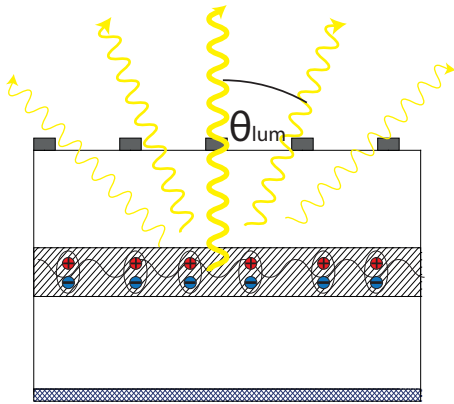}
\caption{Luminescence of a modulated excitonic condensate. Additional rays appear due to the oscillations of the order parameter, with the higher harmonics being suppressed. Thickness of the wavy lines corresponds to the intensity of the radiation. The upper electrode is semitransparent and its effect on the luminescence can be neglected.}
\label{fig:lumin}
\end{figure}

Now we discuss experimental manifestations of anisotropic superfluidity in the proposed realization following the results of Sec.\ref{main}. First of all we note that the density of the condensate is periodically modulated (see (\ref{gpsol}),(\ref{gpsol1})). In the case of direct optical recombination of excitons this will lead to
additional features in their luminescence. For a uniform condensate luminescence is normal to the QW plane\cite{Tbec} with wavevector $k_z$ given by $E_g/\hbar c$,where $E_g$ is the excitonic gap and $c$ is the speed of light in vacuum. In the presence of a modulation $\Phi({\bf r})$ contains harmonics carrying momentum $\pm {\bf q}$. This momentum can be transferred to photons leading to appearance of two additional luminescence rays with momentum $(\pm{\bf q},\sqrt{k_z^2-q^2})$. They will be directed at angles $\theta_{lum}=\pm \arcsin(\left| \hbar {\bf q} \right| c/ E_g)$  with respect to the normal to the QW plane in the cross-sectional plane (see Fig.\ref{fig:lumin}). The intensity of this additional rays will be proportional to $|\Phi_\pm|^2$ (see Eq. \ref{gpsol}). In the case of an external potential consisting of more than one harmonic, the above considerations lead to a "fan" (in 1D case) or a "lattice" of additional luminescence rays (similar effect has been predicted for stimulated many-photon recombination of an exciton BEC in\cite{ssc126000269}). Note that in our model the order parameter should contain higher harmonics; however, their intensity is small. Magnitude of the second harmonic should be $\sim O([V_0/(T+2U)]^2)$ and thus intensity of corresponding luminescence rays is on the order $\sim O([V_0/(T+2U)]^4)$ compared to the central ray.

As a direct consequence of the superfluid density anisotropy, the shape of the angle-resolved luminescence profile close to the normal direction is elongated along ${\bf q}$ and compressed in the perpendicular direction. At finite temperatures the intensity of the quasicondensate luminescence can be calculated by means of hydrodynamic method in quantum field theory
\cite{pit-str,pr0155000080,Popov,qft-hd} with the result being:\cite{angleresolvedlumin}
\begin{equation}\label{Iangle}
I(\vartheta,\varphi)=K\frac{(\tilde cE_g/cT)^{\gamma}}
{[\sin^2\vartheta(1-2\Delta_s\cos^2\varphi)]^{1-\gamma/2}},
\end{equation}
$$
K=\frac{K_0E_gmT\eta}{(2\pi\hbar)^2\tau_{\rm bright}},\;
\gamma=\frac{mT}{2\pi\hbar^2n\sqrt{1-2\Delta_s}},
$$
where $\vartheta$ is the angle between luminescent ray and the normal to the QW plane, $\tilde c=\sqrt{U_0n/m}$, $K_0\sim1$ is a dimensionless constant, $\tau_{\rm bright}$ is the exciton lifetime in the radiative zone, $\eta=N_0/N$ is the zero-temperature condensate fraction and $T$ is the exciton temperature, that is assumed to be finite, but low enough\cite{T>0}. Rays corresponding to higher harmonics of the anisotropic potential acquire analogous anisotropic shape.

Moreover, the luminescence spectrum also acquires an anisotropic form:
\begin{equation}\label{Iomega}
I(\vartheta,\varphi,\omega)=I_{\varphi\vartheta}^0
\delta(E_g+\mu-\varepsilon_{\varphi\vartheta}-\hbar\omega),
\end{equation}
$$
I_{\varphi\vartheta}^0=\frac{K_0E_g^3}{(2\pi\hbar c)^2\tau_{\rm bright}}
\int|v_{\bf o}|^2\frac{d{\bf r}}S,
$$
$$
\varepsilon_{\varphi\vartheta}=(C_s^{\pi/2}/c)E_g
\sin\vartheta\sqrt{1-2\Delta_s\cos^2\varphi},
$$
where $o_x=q_r\sin\vartheta\cos\varphi$, $o_y=q_r\sin\vartheta\sin\varphi$,
$q_r=E_g/\hbar c$, the dependence of $v_{\bf o}$ on ${\bf r}$ is given by
(\ref{bogolsol}), and $C_s^{\pi/2}$ is equal to $C_s$ (see (\ref{csnd})) for $\varphi=\pi/2$. It is remarkable, that the luminescence frequency in (\ref{Iomega}) depends on the in-plane angle $\varphi$. The frequency shift between the directions $\varphi=0$ and $\varphi=\pi/2$ is then given by:
\begin{equation}\label{deltaomega}
\delta\omega=(C_s^{\pi/2}/c)E_g\sin\vartheta(1-\sqrt{1-2\Delta_s}),
\end{equation}
that is evidently non-zero and is determined by the anisotropy parameter $\Delta_s$. A similar effect takes place for the rays corresponding to higher harmonics of the order parameter as well as for the luminescence along the normal to the QW plane if an in-plane magnetic field is applied\cite{prb062001548,prl087216804,ssc144000399}.

\begin{figure}[h!]
\begin{center}
\includegraphics[width=8cm]{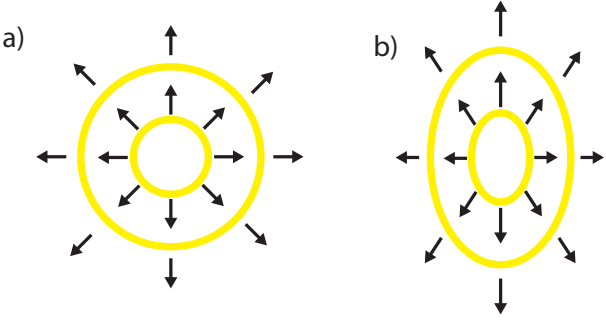}
\end{center}
\caption{Qualitative depiction of the propagation of a a) circular and b) elliptic wave from a point source through isotropic and anisotropic superfluid exciton BEC, correspondingly. Length of the arrows corresponds to the magnitude of the sound velocity in the corresponding direction. The smaller axes of the ellipses are directed along {\bf q}.}
\label{fig:waves}
\end{figure}

Anisotropy of the excitation spectrum (\ref{spectrum}) is at the heart of a number of observable phenomena. First of all, one can directly measure the spectrum experimentally. Techniques for such measurements are known for systems of excitons \cite{prb062001548} and have been described in the literature. Anisotropy of sound velocity (\ref{csnd}) can be investigated by a direct measurement as well \cite{prb062001548}. A different option also exists: as a consequence of the anisotropy of sound, circular waves should become {\it elliptic} with the ratio between axes equal to $C_s^x/C_s^y$ (Fig.\ref{fig:waves}). An elliptical wave can be created by an abrupt change of local chemical potential \cite{prl079000553} caused by a voltage applied to a region of the upper electrode \cite{apl085005830,prl106196806}. The propagation of the wave can be observed then in a time-resolved luminescence experiment\cite{prl094226401}.

Another quantity we are interested in is the helicity modulus (\ref{y_sq}). To determine ${\rm Y}_s(\phi)$ we propose to create 2D excitons by spatially resolved continuous wave pumping, with an in-plane magnetic field being applied in the QW plane during the pump. As it is known \cite{prl087216804}, the presence of crossed out-of-plane electric and in-plane magnetic fields results in a shift of exciton spectrum in the momentum space. The dispersion law takes then the form:
\begin{equation}\label{e0pB}
\varepsilon_0({\bf p}_B)=\frac{({\bf p}_B-{\bf p}_0)^2}{2m},\;\;\;\;
{\bf p}_0\equiv{\bf B}_{\parallel}\times{\bf d}_0/c.
\end{equation}
Here ${\bf p}_B$ is the magnetic momentum, ${\bf p}_0$ is the shift momentum, ${\bf B}_{\parallel}$ is the in-plane magnetic field, and ${\bf d}_0$ is the exciton dipole moment. We also neglect the change in the exciton effective mass due to the magnetic field because we assume ${\bf B}_{\parallel}$ to be small and the correction is quadratic in ${\bf B}_{\parallel}$.

After the collisional relaxation (local thermalization inside the exciton gas) \cite{prb053015834,collrel} and the phonon relaxation (cooling of the locally equilibrated exciton gas) \cite{phonrel} exciton occupancy $n({\bf p}_B)$ "falls" $ $ down to the bottom of the shifted parabola (${\bf p}_B\approx{\bf p}_0$). In this state the exciton system is "cold" and the group velocity of excitons averaged over $n({\bf p}_B)$ is zero\cite{nonzero}:
\begin{equation}\label{v0=0}
{\bf v}_0\equiv
\langle\partial\varepsilon_0({\bf p}_B)/\partial{\bf p}_B\rangle=
\langle{\bf p}_B-{\bf p}_0\rangle/m=0.
\end{equation}
The system is then at rest despite the dispersion law shift.  The cooled excitons flowing to the examined area after their global thermalization and transition into the superfluid state will thus also have a shifted magnetic momentum $\langle{\bf p}_B\rangle={\bf p}_0$.

Suppose now that ${\bf p}_0$ changes with time ${\bf p}_0\to{\bf p}_0-{\bf P}$, where ${\bf P}$ depends on time adiabatically slow.  The normal component will remain at rest due to relaxation processes\cite{10ps}, while the superfluid component will be set into motion. The resulting system velocity will be related to the probe momentum ${\bf P}$ through the helicity modulus tensor (see \ref{j_dfn}). Thus the helicity modulus can be measured.

One way to implement the idea above is to change slowly the polarizing electric field $E_z^{\rm av}\to E_z^{\rm av}+\Delta E_z^{\rm av}(t/\tau_{\rm sw})$, where $\tau_{\rm sw}$ is the characteristic switching time. This results in a change of the exciton dipole moment ${\bf d}_0\to {\bf d}'(t/\tau_{\rm sw})\equiv{\bf d}_0+\Delta{\bf d}(t/\tau_{\rm sw})$ and thus changes the bottom of the shifted parabola (see (\ref{e0pB})), i.e., the quantity ${\bf P}$.

Let us discuss the limitations on the electric field switching time. It is bound from above by the exciton lifetime because in a stationary regime the excitons created by the pump will replace the recombined ones leading to a large number of excitons having momentum lower then the probe one in the system. In contemporary exciton luminescence experiments electric field switching occurs on timescales down to 100 picoseconds \cite{jap104063515} which is guaranteed to be smaller than the usual exciton lifetimes. The lower boundary follows from the fact that in the course of a non-adiabatic perturbation transitions to excited states may occur destroying superfluidity and even heating the system.

\begin{figure}[h!]
\begin{center}
\includegraphics[width=4cm]{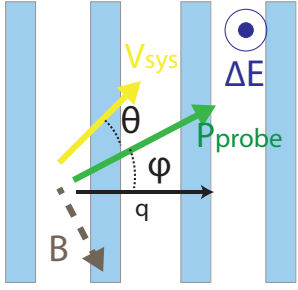}
\end{center}
\caption{Noncollinearity of the non-dissipative flow and the probe momentum in an anisotropic superfluid. $\phi$ is the angle between $P_{probe}$ and the periodic potential orientation, $\theta$ is the angle between the direction of the non-dissipative flow and the probe momentum. The in-plane magnetic field {\bf B} (grey) is oriented perpendicularly to the probe momentum, while the variation of the polarizing electric field $\delta${\bf E} (dark blue) is oriented perpendicularly to the QW plane.}
\label{fig:noncoll}
\end{figure}

As has been discussed in Sec.\ref{anis}, the total current in the system is related to the probe momentum through the helicity modulus tensor and can be noncollinear to it (Fig.\ref{fig:noncoll}). To determine the total current one must know the total density and the velocity of the system's motion. Both quantities can be measured from the recombination luminescence of excitons: the intensity is proportional to the total density and the direction and the magnitude of the velocity can be determined by observing movement of the radiating excitonic spot. Thus knowing the probe momentum from field parameters it is possible to determine the helicity modulus. Note that in sufficiently high magnetic fields exciton recombination is suppressed \cite{prb062001548}; however, phonon-assisted luminescence\cite{prb050001119} should make the observation of exciton motion nevertheless possible.

\begin{figure}[h!]
\begin{center}
\includegraphics[width=8cm]{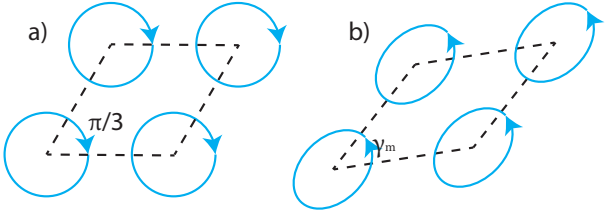}
\end{center}
\caption{Unit cell of the vortex lattice a) without or b) with an external periodic potential. We assume that the periodic potential wavevector is oriented along the small diagonal of the unit cell.}
\label{fig:vortlat}
\end{figure}

There is also a method to measure anisotropy of helicity modulus indirectly. 'Stirring' a condensate with a frequency greater then a critical one is known to lead to formation of quantized Feynman vortices in the system\cite{prl084000806}. Such a 'stirring' can be performed for indirect dipolar excitons by a radial in-plane magnetic field\cite{prl100250401}. In an isotropic case the vortices are expected to form a triangular lattice as a consequence of radially symmetric intervortex interactions [see Fig.\ref{fig:vortlat}a]. In a weak anisotropic case, however, the symmetry of equilateral triangle type is lost [see Fig.\ref{fig:vortlat}b]. The unit cell is  deformed due to an effective rescaling of coordinates\cite{pra086043612,prb044004503} $x\to x/\eta$, $y\to\eta y$,
где $\eta=[({\rm Y}_s)_{xx}/({\rm Y}_s)_{yy}]^{1/4}$. Thus the difference between the minimal angle $\gamma_m$ of the unit cell and $\pi/3$ allows one to measure the anisotropy of the helicity modulus.

\section{\label{estimation} Estimation of the observable effects }
Now we will consider four particular setups, based on the existing structures for exciton condensation observations. Parameters of these structures are given in the upper section of Table \ref{t:struct}. In the two of these setups (GaAs/AlGaAs and MoS$_2$/hBN coupled QWs) applied voltage is uniform and thus $\lambda=a+b$.

\begin{table}
\begin{tabular}{lllll}
&GaAs/&GaAs/&MoS$_2$/&MoSe$_2$/\\
quantity&AlGaAs&AlGaAs&hBN&hBN\\
&GaAs& &MoS$_2$&WSe2\\
&CQWs&SQW&CQWs&CQWs\\
\hline
$E_g$, eV&1.55&1.51&1.8&1.3\\
$L_{\rm QW}$, nm&8&40&0.333&0.333\\
$L_{\rm B}$, nm&4&--&1.667&1\\
$l$, nm&1000&120&11&11\\
$z_0$, nm&100&60&8&8\\
$a$, nm&500&60&6&6\\
$b$, nm&500&70&7&6\\
$\lambda$, nm&1000&130&13&12\\
\hline
$n_0$, $10^{10}$ cm$^{-2}$&0.8&1&80&160\\
$\mu$, meV&0.5&1.1&16.9&27.0\\
$\hbar^2 q^2/2m$, meV&$6.8\cdot10^{-3}$&0.40&8.9&11.9\\
$V_0$, meV&0.15&0.40&8.1&7.8\\
$T_c$, K&0.5&0.48&6.8&23.8\\
\end{tabular}
\caption{\small Upper section: material parameters for the coupled GaAs/AlGaAs QWs \cite{elst}, single GaAs/AlGaAs QW \cite{K,jetpl0940800}, coupled MoS$_2$/hBN QWs \cite{14041418} and coupled MoSe$_2$/hBN/WSe$_2$ QWs \cite{nc0006006242,sci351000688}. Given are the values for the excitonic gap $E_g$, the QW width $L_{\rm QW}$ and the barrier width $L_{\rm B}$, the distances between QW(s) and the bottom electrode $z_0$ and between the top and the bottom electrodes $l$. Lower section: Excitonic parameters for the model (\ref{ham}). Given are the values for the
exciton density $n$, $n\approx n_0$, the chemical potential $\mu$, the amplitude of the external periodic potential $V_0$, the characteristic kinetic energy contribution $\hbar^2 q^2/2m$, and the estimate for $T_c\approx\pi\hbar^2n_s/2m$ for the superfluid transition temperature, where we set \cite{ns} $n_s=n_0\sqrt{1-2\Delta_s}$.}
\label{t:struct}
\end{table}

\begin{figure}[h!]
\begin{center}
\includegraphics[width=\linewidth]{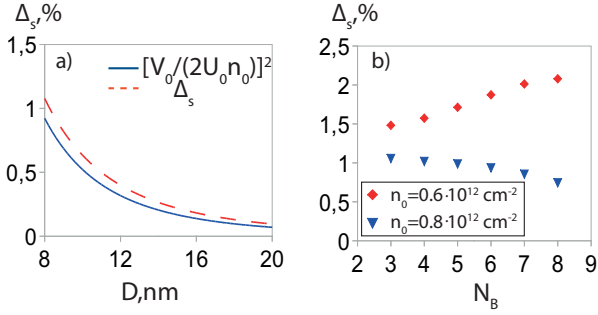}
\end{center}
\caption{Dependence of $\Delta_s$ on the e-h separation distance. a) GaAs/AlGaAs/GaAs CQW structure with $V_0=0.05$ meV. $\Delta_s$ approximately follows $(V_0/(2 U_0 n_0))^2$.
b) MoS2/hBN/MoS2 CQW structure, $V_0=1.4$ meV, for two values of exciton densities $0.6\cdot10^{12}$ cm$^{-2}$ (orange diamonds) and $0.8\cdot10^{12}$ cm$^{-2}$ (blue triangles). Number of hBN barrier monolayers $N_B$ is used instead of $D$. For parameters not given here see Table\ref{t:struct}.}
\label{fig:deltasd}
\end{figure}

We have estimated the magnitude of the predicted effects for experimental setups described in Table \ref{t:struct}. First of all one needs to estimate parameters of the model (\ref{ham}). The interaction between indirect excitons cannot be taken simply as $U_0(r)=(2e^2/\varepsilon)(1/r-1/\sqrt{r^2+D^2})$, because in the dipolar limit $D\to0$, $eD=const$ the Fourier transform of this potential is singular. One has thus to take into account the renormalizations stemming from ladder diagrams and related to the scattering problem\cite{loz-yud}. We use instead a model potential $U({\bf p})$, which is defined as follows. Its 'contact' part $U_0=U(0)$ is deduced from the results of a quantum Monte-Carlo simulation for a system of dipoles without periodic modulation \cite{ssc144000399}. The 'long-range' part of the potential $U({\bf p})-U_0$  is then taken to be the same as for $U_0(r)$(this quantity does not diverge even in the dipolar limit). Details of the estimates are given in Appendix \ref{Ueff}. Actually, the form of the potential (\ref{Up}), (\ref{A4}) leads to interesting qualitative results regarding the anisotropy parameter $\Delta_s$. In Fig.\ref{fig:deltasd}. For the GaAs/AlGaAs/GaAs CQW structure $q$ is small such that $\Delta_s\approx(V_0/(2 U_0 n_0))^2$. Consequently, as $D$ increases, the dipole-dipole interactions become stronger and $\Delta_s$ is strongly suppressed. For MoS2/hBN/MoS2 CQW structure, on the contrary, an increase or a very slow decrease of $\Delta_s$ can be observed. The explanation is that for this case the position of a rotonic-like minimum of the function $k^2/2m+2 U({\bf k}) n_0$ is very close to ${\bf q}$. In strongly correlated systems the position of the rotonic minimum is given by $2\pi\hbar\sqrt{n}$\cite{ssc144000399,prl098060405} which nearly coincides with $q=2\pi\hbar/\lambda$ for density $0.6\cdot10^{12}$ cm$^{-2}$ ($\lambda=13$ nm from Table \ref{t:struct}, $1/\sqrt{n}=12.9$ nm). Thus, for this case $T+2U$ is expected to decrease when interactions become stronger, until an instability is reached\cite{prb090165430}. Here we restrict our considerations to systems without rotonic instability and we have checked that for the parameters used in Table \ref{t:struct} the excitation spectrum (\ref{spectrum}) is stable.

Let us move to calculations for realistic system parameters. Calculation of the electric field configuration in the QW plane (including estimates for $E_z^{\rm av}$ and $V_0$) is presented in Appendix \ref{dissoc}. Single-exciton properties, such as the electron-hole separation, the lifetime and the binding energy have been obtained from a numerical solution of the Schr\"odinger equation. This is discussed in detail in Appendix \ref{1DSE&EB}.

We turn now to the helicity modulus measurement procedure described in Sec.\ref{phys}. If the direction of the probe momentum constitutes an angle $\phi$ with the $x$ axis the angle $\theta$ between system's velocity ${\bf v}_{\rm sys}$ and probe momentum (see Fig.\ref{fig:noncoll}) is:
\begin{equation}\label{theta}
\theta=
\arccos\frac{1-2\Delta_s\cos^2\phi}
{\sqrt{1-4\Delta_s\cos^2\phi+4\Delta_s^2\cos^2\phi}},
\end{equation}
where $\Delta_s$ is the anisotropy parameter defined in (\ref{Deltas}). This angle is maximal at a certain $\phi=\phi_m$ and has value $\theta_m$:
\begin{equation}\label{phimthetam}
\phi_m=\arccos\frac1{\sqrt{2(1-\Delta_s)}},\;\;\;\;
\theta_m=\arccos\frac{\sqrt{1-2\Delta_s}}{1-\Delta_s}.
\end{equation}
Moreover, we have estimated the velocity acquired by the system after the electric field switching procedure discussed in Sec.\ref{phys}. For GaAs structures it is $3.2\cdot 10^5$ cm/sec for coupled QWs and $9.6\cdot10^5$ for single QW in magnetic field $B_\parallel=8$ T (see details in Appendix \ref{accel}). However, for the other two structures this method has turned out to be unfeasible. Alternative ways are to create a gradient of a local chemical potential of excitons\cite{grad} or to move a macroscopically coherent exciton system along a narrow channel \cite{narrowchannel}.

Now we move onto the indirect effects discussed in Sec.\ref{phys}. Their magnitude can also be related to $\Delta_s$. Ratio of the axes of an elliptical wave is $C_s^x/C_s^y \approx 1- \Delta_s$. Thus a good measure of anisotropy is the quantity $\delta C_s/C_s=1-C_s^x/C_s^y$. Calculation of the minimal angle in the deformed vortex lattice unit cell is also straightforward for the case when the period of the vortex lattice is much larger than $\lambda$, as one can simply rescale the parameters of the unit cell. We assume that the principal axes of the helicity modulus tensor are along the diagonals of the unit cell (which is a rhombus). If $x$ axis is along the larger diagonal then it is contracted by a factor of  $\sqrt{1-2\Delta_s}$. It follows then that the minimal angle in the unit cell $\gamma_m$ is:
\begin{equation}\label{gm}
\gamma_m=\arccos{\frac{1+\Delta_s}{2-\Delta_s}}.
\end{equation}
A measure of the anisotropy of quasicondensate luminescence intensity for directions close to normal to the QW plane is given by:
\begin{equation}\label{deltaI}
\delta I/I=[I(\vartheta,0)-I(\vartheta,\pi/2)]/I(\vartheta,0)=2\Delta_s.
\end{equation}
Corresponding frequency shift between the luminescence along and across ${\bf q}$ is given by (\ref{deltaomega}).

All of the results of estimations discussed above are summarized in Table \ref{t:anissf}. One can see that the anisotropy effects are weak for large $\lambda$. However if  $\lambda$ becomes of the order of the interexciton distance, the effects are considerably enhanced, so that an intermediate anisotropy regime $({\rm Y}_s)_{yy}/({\rm Y}_s)_{xx}\sim3$ is realized.
\begin{table}
\begin{tabular}{lllll}
&GaAs/&GaAs/&MoS$_2$/&MoSe$_2$/\\
quantity&AlGaAs&AlGaAs&hBN&hBN\\
&GaAs& &MoS$_2$&WSe2\\
&CQWs&SQW&CQWs&CQWs\\
\hline
$\Delta_s$&2.9\%&21.4\%&31.1\%&5.5\%\\
$\phi_m$&$44^{\circ}$&$37^{\circ}$&$32^{\circ}$&$43^{\circ}$\\
$\theta_m$&$1.7^{\circ}$&$15.8^{\circ}$&$26.9^{\circ}$&$3.3^{\circ}$\\
$v_{\rm sys}/v$&0.97&0.76&0.61&0.94\\
$\delta C_s/C_s$&3\%&24\%&39\%&6\%\\
$\gamma_m$&$58.6^{\circ}$&$47.1^{\circ}$&$39.1^{\circ}$&$57.2^{\circ}$\\
$({\rm Y}_s)_{yy}/({\rm Y}_s)_{xx}$& 1.06&1.8&2.6&1.1\\
$\delta I/I$&5.7\%&42.9\%&62.2\%&10.9\%\\
$\delta \omega$, $\mu$eV&2.9&35.8&126.9&17.4
\end{tabular}
\caption{\small Estimates for the predicted effects for structures described in Table \ref{t:struct}. Given are the values for the anisotropy parameter $\Delta_s$, the angle of motion of superfluid component $\phi_m$ and the anisotropy angle for the superfluid flow $\theta_m$ (\ref{phimthetam}), the anisotropy of the sound velocity, $\delta C_s/C_s$ , the lowest angle in the triangular unit cell of vortex lattice $\gamma_m$ (\ref{gm}), the ratio of the diagonal components of the helicity modulus tensor $({\rm Y}_s)_{yy}/({\rm Y}_s)_{xx}$ and the degree of the quasicondensate luminescence intensity anisotropy $\delta I/I$ (\ref{deltaI}) as well as frequency anisotropy $\delta \omega=\omega(\vartheta,0)-\omega(\vartheta,\pi/2)$ for $\vartheta=5\pi/12$ (see (\ref{deltaomega})).}
\label{t:anissf}
\end{table}

In Sec.\ref{phys} we have also discussed the limitations which our system should satisfy. In Appendix \ref{app2} a detailed discussion of these limitations is presented with the conclusions that setups considered here do satisfy the necessary conditions.

\section{\label{fin} Conclusion}
In the Article, we have demonstrated anisotropy of helicity modulus, sound velocity and angle-resolved luminescence spectrum for a moving two-dimensional gas of weakly interacting bosons in a one-dimensional external periodic potential. Analytical expressions for anisotropic corrections to the excitation spectrum (\ref{spectrum}), sound velocity (\ref{csnd}) and helicity modulus (\ref{y_sq}),(\ref{Ysmatr}) have been obtained with Bogoliubov technique at $T=0$. An expression for angle-resolved photoluminescence intensity(\ref{Iangle}) has been obtained at low temperatures by means of quantum-field hydrodynamics. The considered model can be used to describe a physical system of dipolar excitons in a QW in an electrostatic lattice. Our calculations can be also applied to systems of dipolar atoms in optical lattices in periodic fields. Our results can be straightforwardly generalized for more complicated forms of periodic potentials as well as systems with intrinsic anisotropy of mass (\ref{YsmatrAnisMass}). We have not taken exciton spin into account, as in the considered regime (see Sec.\ref{main}) these can be neglected.

We propose several qualitative manifestations of excitonic anisotropic superfluidity:

$\bullet$) The photoluminescence of the excitonic system is organized into a pattern of discrete rays with intensity decreasing away from the normal to the QW plane (see Fig. \ref{fig:lumin}). At finite temperatures, due to luminescence of a 2D quasicondensate of excitons each ray has a finite angular extent and an elliptic, rather then circular, shape. This effect is directly related to the anisotropy of the helicity modulus (see Eq.(\ref{Iangle})).

$\bullet$) The unit cell of the triangular vortex lattice, appearing in a radial magnetic field \cite{prl100250401} in the QW plane, will not be equilateral.

$\bullet$) Collisionless sound waves, created by a point-like source will be elliptical instead of circular.

$\bullet$) The momentum transferred to the system will not be collinear to the resulting non-dissipative current.

$\bullet$) The frequency of the angle-resolved luminescence arising from the non-condensate excitons depends on the in-plane direction of the beam (i.e. polar angle $\varphi$). Moreover, if an in-plane magnetic field is applied, the luminescence frequency along the normal to the QW plane depends on the direction of the field.

We have also proposed an experiment to determine the helicity modulus tensor including a method for setting dipolar particles into motion which is valid for other realizations such as atomic systems. Using the results of simulations \cite{ssc144000399} estimates for the magnitude of the predicted effects and manifestations of anisotropic superfluidity have been given. For one of the considered structures we have observed an increase in anisotropy due to closeness of the position of a rotonic-minimum-like feature in the interexciton potential $U({\bf p})$ to ${\bf q}$ (Fig.\ref{fig:deltasd}). The magnitudes of anisotropic effects in Table\ref{t:anissf} give evidence for possibility of their observation and detection in GaAs/AlGaAs heterostructures as well as MoS$_2$/hBN/MoS$_2$ and MoSe$_2$/hBN/WSe$_2$ bilayers in future experiments.

The work was supported by grant Russ. Sci. Found. 17-12-01393.

\appendix

\section{Details of calculations} \label{app1}
\subsection{Calculation of the exciton-exciton interaction potential}\label{Ueff}
Neglecting fermionic and spin effects for the excitons, one can write the Fourier transform $U_{\bf p}$ of the pseudopotential $U({\bf r})$ of the exciton-exciton interaction as:
\begin{equation}\label{Up0}
U_{\bf p}=U_0+\int d{\bf r}(e^{-\frac i{\hbar}{\bf pr}}-1)U({\bf r}),
\end{equation}
where $U_0=\left.U_{\bf p}\right|_{{\bf p}=0}$. In the second term in (\ref{Up0}) we substitute $U({\bf r})$ with the bare interexciton interaction in an $e$-$h$ bilayer:
\begin{equation}\label{Ur0}
U_0({\bf r})=\frac{2e^2}{\varepsilon}
\left(\frac1r-\frac1{\sqrt{r^2+D^2}}\right).
\end{equation}
As a result (\ref{Up0}) takes the form:
\begin{equation}\label{Up}
U_{\bf p}=U_0+\frac{4\pi d^2}{\varepsilon D}
\left(\frac{1-e^{-pD/\hbar}}{pD/\hbar}-1\right),
\end{equation}
where $d=eD$.

We cannot, however, simply use $U_0=\int U_0({\bf r})d{\bf r}$ to calculate $U_0$ in (\ref{Up}), because this function shows a diverging behavior for dipolar interactions ($U_0(r)=d^2/r^3$ has an unintegrable singularity at ${\bf r}=0$). Instead we use the results of an {\it ab initio} modeling \cite{ssc144000399} performed for dipolar excitons.
\begin{equation}
U_0=\frac{\hbar^2}m\frac{\partial^2}{\partial\bar n^2}\bar ne_0(\bar n).
\label{A4}
\end{equation}
Here $\bar n=nm^2d^4/(\hbar^4\varepsilon^2)$ --- dimensionless density and
\begin{equation}
e_0(\bar n)=
a_e\exp(b_e\ln\bar n+c_e\ln^2\bar n+d_e\ln^3\bar n+e_e\ln^4\bar n),
\label{A5}
\end{equation}
--- dimensionless ground state energy per particle, where coefficients $a_e=9.218$, $b_e=1.35999$, $c_e=0.011225$, $d_e=-0.00036$ and $e_e=-0.0000281$ correspond to an interval $1/256\le\bar n\le8$. For all numerical estimates we replace $n$ by $n_0$ in (\ref{A4}), (\ref{A5}) due to the condition $(N-N_0)/N\ll1$ (see Sec. \ref{main}).

\subsection{ Electric Field Distribution in QW Plane}\label{dissoc}
Electrostatic field configuration in the QW plane can be calculated analytically: neglecting inhomogeneities in the charge distribution over the stripes of the upper electrode the problem is solved by image method with respect to the bottom electrode plane (see setup in Fig. \ref{fig:exp}). Assuming the thickness of stripes to be small and denoting charge of a stripe per unit area as $\sigma$ we have:
\begin{equation}
\begin{gathered}
E_x (x,z) = \frac{\sigma}{\varepsilon}
\sum_{j\in Z} \left[
    \ln \left(
    \frac{
    \left[ (x-a-j\lambda )/l\right]^2 + \left[1+z/l\right]^2
    }
    {
    \left[(x-j\lambda )/l\right]^2 + \left[1+z/l\right]^2
    }
    \right)
\right.
\\
\left.
    -\ln \left(
    \frac{
    \left[(x-a-j\lambda )/l\right]^2 + \left[1-z/l\right]^2
    }
    {
    \left[(x-j\lambda )/l\right]^2 + \left[1-z/l\right]^2
    }
    \right)
\right],
\\
\\
E_y(x,z)=0,
\\
\\
E_z (x,z) = \frac{2 \sigma}{\varepsilon}
\sum_{j \in Z} \text{\Huge $[$}
\\
\left.
\arctan \left(
    \frac{
    a/l
    }
    {
    \left[1-z/l\right] \left[(x-j\lambda -a)(x-j\lambda )/(l-z)^2+1\right]
    }
    \right)
\right.
\\
\left.
    +\arctan \left(
    \frac{
    a/l
    }
    {
    \left[1+z/l\right] \left[(x-j\lambda -a)(x-j\lambda )/(l+z)^2+1\right]
    }
    \right)
\right],
\end{gathered}
\label{electrostatic}
\end{equation}
where $\varepsilon$ is the dielectric constant and $\sigma=\varepsilon E_z^{\rm av}\lambda/4\pi a$.

\begin{figure}[h!]
\begin{center}
\includegraphics[width=8cm]{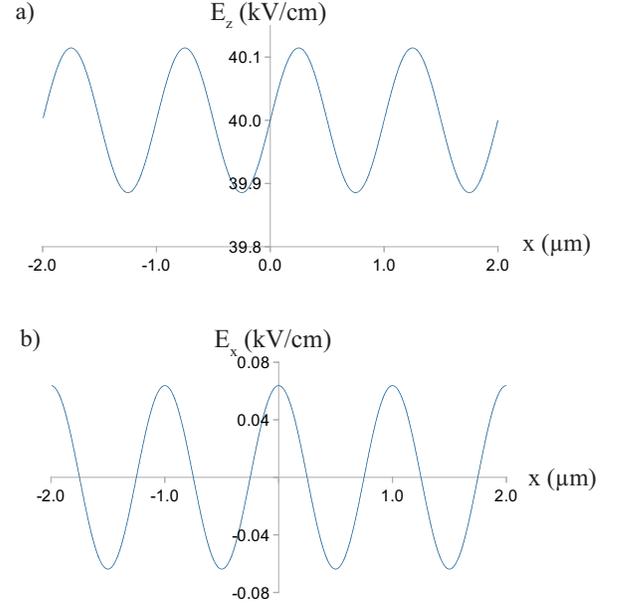}
\end{center}
\caption{Coordinate dependence of the electric field in the QW plane: a) Component normal to the QW plane b) In-plane component}
\label{fig:efield}
\end{figure}

We calculated the field configurations for four setups (see Table \ref{t:struct}) with $E^{av}_z$ given in Table \ref{t:efield}. Summation in (\ref{electrostatic}) was carried out numerically with relative error estimate $10^{-13}$. The result for the first structure is presented in Fig. \ref{fig:efield}. One can see that the oscillations of the electric field have a well defined period equal to $a+b$. This means that if we decompose $E_z(x,z_0)=E_z^{(0)} + \Delta E_z \cos(2 \pi (x-x_0) / \lambda) + E_z^{(2)} \cos(4 \pi(x-x_0) / \lambda) + ...$ then $E_z^{(n)}\ll \Delta E_z$. We verified this by numerical convolution with higher harmonics. For all structures we found that the component along $z$ has a constant component $E_z^{\rm av}$ and an oscillating component with amplitude $\Delta E_z$, while the field along $x$ is purely oscillatory with amplitude $\Delta E_x$ (numerical values presented in Table \ref{t:efield}).

\begin{table}
\begin{tabular}{lllll}
&GaAs/&GaAs/&MoS$_2$/&MoSe$_2$/\\
quantity&AlGaAs&AlGaAs&hBN&hBN\\
&GaAs& &MoS$_2$&WSe2\\
&CQWs&SQW&CQWs&CQWs\\
\hline
$E_z^{\rm av}$, kV/cm&40&5.8&256&447\\
$\Delta E_z$, kV/cm&0.11&0.22&40&58\\
$\Delta E_x$, kV/cm&0.06&0.22&41.8&60.0
\end{tabular}
\caption{\small Electrostatic field parameters in the QW plane for the structures in Table \ref{t:struct}. $E_z^{\rm av}$ is the constant component of the field, $\Delta E_z$ and $\Delta E_x$ are the amplitudes of first harmonic along z and x}
\label{t:efield}
\end{table}
The role of the constant component $E_z^{\rm av}$ is to fix the dipole moment of the excitons while $\Delta E_z$ determines $V_0$ in the model (\ref{ham}). Component $E_x(x,z)$ is oriented in the QW plane and can cause, as has been discussed above, dissociation of the excitons. However, if the energy associated with this field is less then the binding energy of an exciton, dissociation is forbidden. For corresponding estimates see Appendix \ref{app2}.

\subsection{Calculation of the electron-hole separation, the exciton lifetime,
binding energy, and radius}\label{1DSE&EB}
Electron-hole separation is given by:
$$
D=\left|\int(\psi_h^2(z)-\psi_e^2(z))zdz\right|,
$$
where $\psi_{e,h}(z)$ are electron (hole) ground-state wavefunctions which are
satisfied the following 1D Schr\"odinger equation
\begin{equation}\label{SE}
\left(-\frac{\hbar^2}{2m_{e,h}^z}\frac{d^2}{dz^2}+
U_{e,h}(z)\pm eE_z^{\rm av} z-E_0^{e,h}\right)\psi_{e,h}(z)=0.
\end{equation}
and normalized according to $\int\psi^2_{e,h}(z)dz=1$. In Eq.~(\ref{SE}) $e>0$
is the absolute value of the particle charge ("+" is for electron and "-" for
hole), $z$ axis is along the normal to QW plane, $U_{e,h}(z)$ is the QW
potential for electron (hole), $m_{e(h)}^z$ and $E_0^{e(h)}$ are effective
masses along $z$ axis and ground state energies for electron (hole),
respectively.\cite{notunnel} Parameters of the QW structures used
to solve (\ref{SE}) are given in Tab.~\ref{t:1DSE}

\begin{table}
\begin{tabular}{llll}
&GaAs/&GaAs/&MoS$_2$/\\
quantity &AlGaAs&AlGaAs&hBN\\
&CQWs&SQW&CQWs\\
\hline
$m/m_0$&0.22&0.22&1\\
$m_e/m_0$&0.067&0.067&0.5\\
$\varepsilon$&12.5&12.5&6.7\\
$m_e^w/m_0$&0.067&0.067&0.5\\
$m_e^b/m_0$&0.067&0.067&0.5\\
$m_h^w/m_0$&0.4&0.4&0.5\\
$m_h^w/m_0$&0.4&0.4&0.5\\
$U_e^0$, eV&0.3&0.3&--\\
$U_h^0$, eV&0.15&0.15&--\\
$U_e^0-E_0^e$, eV&--&--&3\footnote{\label{fn1} Ref.~\cite{14041418}.}\\
$U_h^0-E_0^h$, eV&--&--&3$^\text{\ref{fn1}}$\\
$\tau_{\rm dir}$, ps&100\footnote{Ref.\cite{prb059001625}.}&%
200\footnote{Ref.\cite{K}.}&0.4 \footnote{Ref.\cite{prb088075434}.}\\
$\tau_{\rm bright}$, ns&150&2&100
\end{tabular}
\caption{\small
Parameters used to solve the 1D Schr\"odinger equation along $z$: the exciton
effective mass $m$, the in-plane electron mass $m_e$, the dielectric constant
$\varepsilon$, the transversal effective masses for the electron (hole) in the
QW $m_e^w$ ($m_h^w$) and in the barrier $m_e^b$ ($m_h^b$), the corresponding
barrier potential magnitude $U_e^0$ and $U_h^0$ for GaAs-based structures, the
tunneling barrier energy $U_e^0-E_0^e$ ($U_h^0-E_0^h$) for MoS$_2$/hBN
structure, the radiative lifetimes of an exciton $\tau_{\rm dir}$in an electric field ${\bf E}=0$ and an indirect exciton $\tau_{\rm bright}$ in the radiative zone for ${\bf E}=\{0,0,E_z^{\rm av}\}$.}
\label{t:1DSE}
\end{table}

Exciton lifetime $\tau$ is estimated to be $\tau\sim50\;\tau_{\rm bright}$ for
GaAs-based structures and $\tau\sim\tau_{\rm bright}$ for the MoS$_2$/hBN
structure \cite{tau}, as
\begin{equation}\label{taubright}
\tau_{\rm bright}=M^2\tau_{\rm dir},\;M\equiv\int\psi_e(z)\psi_h(z)dz,
\end{equation}
where $\tau_{\rm dir}$ is the lifetime of a direct exciton in zero electric field.

Binding energy and average in-plane electron-hole separation are calculated as $E_B=-H(\Delta x_m)$
and $r_{\rm ex}=\Delta x_m$, respectively (see also variational calculation
results \cite{FTT}). Here \cite{uncertrel}
\begin{equation}\label{HDx}
H(\Delta x)=\frac{\hbar^2}{2m_{\rm r}\Delta x^2}-
\frac{e^2/\varepsilon}{\sqrt{D^2+\Delta x^2}},
\end{equation}
$m_{\rm r}=m_e(m-m_e)/m$ is the reduced mass of $e$ and $h$, and $\Delta x_m$
corresponds to the minimum of function (\ref{HDx}).

We estimate the effective exciton diameter due to internal electron-hole structure as twice the average distance between the center of mass and the position of the lighter carrier:
\begin{equation}\label{aex}
a_{\rm ex}=r_{\rm ex}(1+\sqrt{1-4m_{\rm r}/m}).
\end{equation}
The exciton core diameter arising from dipole-dipole interactions between excitons is given by the s-wave scattering length. To improve the accuracy, we use an energy-dependent\cite{pra081013612} s-wave
scattering length\cite{jetp10600296}:
\begin{equation}\label{asdd}
a_s^{\rm dd}=a_s^a\exp(b_s^a\ln p+c_s^a\ln^2p+d_s^a\ln^3p),
\end{equation}
where $p\sim\sqrt{2E/N}$, $a_s^a=0.68845$, $b_s^a=-0.45897$, $c_s^a=-0.03098$ and $d_s^a=0.002096$. For the considered regime $a_{\rm ex}\lessapprox a_s^{\rm dd}$ (see Table \ref{t:anal}) the real exciton diameter is given by $a_s^{\rm dd}$ rather than $a_{\rm ex}$.

\subsection{Acceleration of the condensate with electric field switching}\label{accel}
We have calculated the estimates for the velocity acquired by excitons set into motion with the procedure described in Sec.\ref{phys}. Results are presented in Table \ref{t:move}.
\begin{table}
\begin{tabular}{llll}
&GaAs/&GaAs/\\
quantity&AlGaAs&AlGaAs\\
&CQWs&SQW\\
\hline
$B_{\parallel}$, T&8&8\\
$\Delta D$, nm&0.5&1.5\\
$v$, cm/sec&$3.2\cdot10^5$&$9.6\cdot10^5$\\
excitons are&dark&dark
\end{tabular}
\caption{\small Parameters for the proposed method of the condensate acceleration: the in-plane magnetic field $B_{\parallel}$, the magnitude of the e-h separation change $\Delta D$ and the resulting system velocity $v$.}
\label{t:move}
\end{table}
We note that the switching is fast enough to ignore the exciton recombination, but slow enough to be considered adiabatic and ignore the normal component:
\begin{itemize}
\item Exciton lifetime $\tau$ (see Table \ref{t:anal}) is much larger than the switching time $\tau_{\rm sw}$. Consequently, exciton recombination does not affect velocity of superfluid motion.

\item On the other hand, $\tau_{\rm sw}$ is much larger then $\tau_{\rm dissip}^{\rm norm}$ --- time of normal component dissipation \cite{10ps}. Thus the normal component is approximately at rest during switching.

\item In the contemporary experiments \cite{Bbec,exexp} on exciton BEC the characteristic size $L$ of the system is of the order of $10-100\;\mu$m and is much smaller than $L_{\rm max}^{\rm adiab}\equiv\pi C_s\tau_{\rm sw}$. It follows then that the switching process does not noticeably excite the system and thus can be considered as adiabatical.
\end{itemize}

\section{Analysis of experimental realization of the effects}\label{app2}
Feasibility of the proposed experiments is supported by the data summarized in Table \ref{t:anal}.

\begin{table}
\begin{tabular}{lllll}
&GaAs/&GaAs/&MoS$_2$/&MoSe$_2$/\\
quantity&AlGaAs&AlGaAs&hBN&hBN\\
&GaAs& &MoS$_2$&WSe2\\
&CQWs&SQW&CQWs&CQWs\\
\hline
$\tau$, $\mu$s& $>$6&$>$0.1&$>$0.1&$>$0.1\footnote{Ref.\cite{sci335000947}}\\
$\tau_{\rm sw}$, ns&8&5&--&--\\
$\tau_{\rm dissip}^{\rm norm}$, ps&10&10&--&--\\
$L_{\rm max}^{\rm adiab}$, $\mu$m&501&473&--&--\\
$C_s$, $10^5$ cm/sec&20&30&57&74\\
$c_{\rm LA}^{\rm semic}$, $10^5$ cm/sec&5.36&5.36&7.11\footnote{Ref. \cite{srep05722}}&4.1\footnote{Ref. \cite{prb090045422}}\\
$E_{\rm dissoc}^{\rm in-plane}$, meV&2.0&0.9&17&23\\
$r_{\rm ex}$, nm&22&26&2.1&2.3\\
$a_{\rm ex}$, nm&30.7&36.1&2.7&2.3\\
$a_s^{\rm dd}$, nm&30.0&34.3&3.7&2.2\\
$E_B$, meV	&2.8&2.4&43.1&50.6\\
$E_z^{\rm ind-dir}$, kV/cm&7\footnote{Ref.\cite{prb059001625}.}&4\footnote{Ref.\cite{K}.}&$\sim2000$\footnote{Ref.~\cite{14041418}.}&$\sim2000$\footnote{Ref.~\cite{nc0006006242}.}\\
\end{tabular}
\caption{\small Parameters of the proposed realizations: the exciton lifetime $\tau$, switching time for the electric field (see Sec.\ref{phys}) $\tau_{\rm sw}$, the normal component dissipation time $\tau_{\rm dissip}^{\rm norm}$, the maximal system size for the electric field switching to be adiabatic $L_{\rm max}^{\rm adiab}$, the sound velocity in the exciton superfluid $C_s$ (average value neglecting periodic potential), the sound velocity for longitudinal acoustic phonons in a semiconductor $c_{\rm LA}^{\rm semic}$, the dissociation energy by the in-plane field $E_{\rm dissoc}^{\rm in-plane}$, the average in-plane electron-hole separation, $r_{\rm ex}$, the effective exciton diameter $a_{\rm ex}$, the energy-dependent s-wave scattering length due to dipolar interactions $a_s^{\rm dd}$, the exciton binding energy $E_B$, and the maximal electric field for which the transition of spatially indirect excitons into direct ones is allowed $E_z^{\rm ind-dir}$. Structures are the same as in Table \ref{t:struct}.}
\label{t:anal}
\end{table}

\begin{itemize}
\item $C_s$ for the excitonic system is larger than the sound velocity $c_{\rm LA}^{\rm phon}$ for longitudinal acoustic phonons in the QW material. This enables efficient cooling of excitons by semiconductor lattice through emission of "Cherenkov" phonons. Because of this excitons can cool down to temperatures as low as $T=0.1$K \cite{Bbec} during their lifetime. This temperature is evidently smaller than the estimate for superfluid crossover $T_c$.

\item Binding energy of an exciton $E_B$ is larger than the sum of the chemical potential $\mu$ and the dissociation energy due to an in-plane field $E_{\rm dissoc}^{\rm in-plane}=eE_x^0\lambda/\pi$. This means that the dissociation of an exciton by tunneling of e and h to neighboring nodes of the in-plane field $E_x(x)=E_x^0\sin(2\pi x/\lambda)$ (which is most profitable energetically) is forbidden.

\item Effective exciton diameter $a_{\rm ex}$ is close to (or smaller then) the energy-dependent s-wave scattering length $a_s^{\rm dd}$ due to the dipole-dipole interactions. It follows then that the overlap between the wavefunctions of the neighboring excitons is not too large and the exchange effects can be neglected at least for qualitative purposes (i.e. fermionic effects are not too important and the excitons can be considered as bosons).

\item In GaAs coupled QWs transformation of spatially indirect excitons into direct ones does not take place. The reason is that the maximal electric field $E_z^{\rm ind-dir}$ for which this is possible is smaller then the minimal value of $E_z$. In MoSe$_2$/hBN/WSe$_2$ QWs an exciton ground state corresponds to an indirect exciton. Therefore, since the maximal value of $E_z$ is smaller than $E_z^{\rm ind-dir}$, the indirect -- direct exciton transition is forbidden as well. In MoS$_2$/hBN/MoS$_2$ QWs, on the contrary, for the parameters considered this transition is allowed. The transition is nonresonant and must be accompanied by emission of a phonon. This gives an additional nonradiative channel of indirect exciton decay with characteristic time set by scattering on acoustic phonons. In the case we have considered it will be suppressed due to the relatively large interwell distance.

\item According to the results of our calculation in MoS$_2$/hBN QW for electron-hole separation $D=2$ nm
$E_z^{\rm av}=256$ kV/cm compensates the $z$-component of $eE_z^{eh}(D)=\left.d(\mu(z)-E_B(z))/dz\right|_{z=D}$ --- the electron-hole attraction in an indirect exciton. In this case disorder caused by fluctuations of hBN barrier width is suppressed which is important for superfluid properties \cite{pit-str}. For MoSe$_2$/hBN/WSe$_2$ QWs, which have $D=1.333$ nm, the field $E_z^{\rm av}=447$ kV/cm also corresponds to compensation.
\end{itemize}


\begin{thebibliography}{99}
\bibitem{BKT}
V. L. Berezinskii, JETP {\bf 32}, 493 (1970); {\it ibid.} {\bf 34}, 610
(1971); J. M. Kosterlitz and D. J. Thouless, J. Phys. C {\bf 6}, 1181 (1973);
J. M. Kosterlitz, {\it ibid.} {\bf 7}, 1046 (1974); D. R. Nelson and J. M.
Kosterlitz, Phys. Rev. Lett. {\bf 39}, 1201 (1977).
\bibitem{atombec}
K.B. Davis, M.O. Mewes, M.R. Andrews, N.J. van Druten, D.S. Durfee, D.M. Kurn, and W. Ketterle, Phys. Rev. Lett. {\bf 75}, 3969 (1995);
 M.H. Anderson, J.R. Ensher, M.R. Matthews, C.E. Wieman, and E.A. Cornell, Science, {\bf 269}, 198 (1995).
\bibitem{prl091250401}M. W. Zwierlein, C. A. Stan, C. H. Schunck, S. M. F. Raupach, S. Gupta, Z. Hadzibabic, and W. Ketterle, Phys. Rev. Lett. {\bf 91}, 250401 (2003).
\bibitem{nature441001118} Z. Hadzibabic, P. Kr\"uger, M. Cheneau, B. Battelier, and J. Dalibard, Nature {\bf 441},1118 (2006).
\bibitem{pitstrbook} L. Pitaevskii, S. Stringari, Bose-Einstein Condensation,  ISBN-13: 978-0198507192

\bibitem{Bbec}
A. A. High, J. R. Leonard, A. T. Hammack, M. M. Fogler, L. V. Butov, A. V.
Kavokin, K. L. Campman, and A. C. Gossard, Nature {\bf 483}, 584 (2012); A. A.
High, J. R. Leonard, M. Remeika, L. V. Butov, M. Hanson, and A. C. Gossard,
Nano Lett. {\bf 12}, 2605 (2012); A. A. High, A. T. Hammack, J. R. Leonard, S.
Yang, L. V. Butov, T. Ostatnick\'y, M. Vladimirova, A. V. Kavokin, T. C. H.
Liew, K. L. Campman, and A. C. Gossard, Phys. Rev. Lett. {\bf 110}, 246403
(2013).

\bibitem{nature443000409}
J. Kasprzak, M. Richard, S. Kundermann, A. Baas, P. Jeambrun, J. M. J.
Keeling, F. M. Marchetti, M. H. Szyma$\acute{n}$ska, R. Andr$\acute{e}$, J. L.
Staehli, V. Savona, P. B. Littlewood, B. Deveaud and Le Si Dang, Nature {\bf
443}, 409 (2006).

\bibitem{LY}
Yu. E. Lozovik, V.I. Yudson, JETP Lett. {\bf 22}, 274 (1975); JETP {\bf 44},
389 (1976); Solid State Commun. {\bf 19}, 391 (1976); {\it ibid.} {\bf 21},
211 (1977); {\it ibid.} {\bf 22}, 117 (1977).

\bibitem{2dde}
L. V. Butov, J. Phys.: Condens. Matter {\bf 16}, R1577 (2004); V. B. Timofeev,
A. V. Gorbunov and A. V. Larionov, {\it ibid.} {\bf 19}, 295209 (2007); R.
Rapaport and G. Chen, {\it ibid.} {\bf 19}, 295207 (2007); M. Combescot, O.
Betbeder-Matibet, and F. Dubin, Phys. Rep. {\bf 463}, 215 (2008); D. W. Snoke,
Adv. Cond. Matt. Phys. {\bf 2011}, 938609 (2011); L. V. Butov, JETP {\bf 122},
434 (2016).

\bibitem{K}
V. V. Solov'ev, I. V. Kukushkin, J. Smet, K. von Klitzing, and W. Dietsche,
JETP Lett. {\bf 83}, 553 (2006); {\it ibid.} {\bf 84}, 222 (2006).

\bibitem{F}
A. Filinov, P. Ludwig, Yu. E. Lozovik, M. Bonitz, and H. Stolz, J. Phys.:
Conf. Series {\bf 35}, 197 (2006); P. Ludwig, A. Filinov, M. Bonitz, and H.
Stolz, Phys. Stat. Solidi B {\bf 243}, 2363 (2006); K. Sperlich, P. Ludwig, A.
Filinov, M. Bonitz, H. Stolz, D. Hommel, and A. Gust, Phys. Stat. Solidi C
{\bf 6}, 551 (2009).

\bibitem{prb042007434}
T. C. Damen, J. Shah, D. Y. Oberli, D. S. Chemla, J. E. Cunningham, and J. M.
Kuo, Phys. Rev. B {\bf 42}, 7434 (1990).



\bibitem{prb059001625}
L. V. Butov, A. Imamoglu, A. V. Mintsev, K. L. Campman, and A. C. Gossard,
Phys. Rev. B {\bf 59}, 1625 (1999).

\bibitem{prl086005608}
L. V. Butov, A. L. Ivanov, A. Imamoglu, P. B. Littlewood, A. A. Shashkin, V.
T. Dolgopolov, K. L. Campman, and A. C. Gossard, Phys. Rev. Lett. {\bf 86},
5608 (2001).

\bibitem{prb053015834}
C. Piermarocchi, F. Tassone, V. Savona, A. Quattropani, and P. Schwendimann,
Phys. Rev. B {\bf 53}, 15834 (1996).

\bibitem{nl0007001349}
A. G. Winbow, A. T. Hammack, L. V. Butov, and A. C. Gossard, Nano Lett. {\bf
7}, 1349 (2007).

\bibitem{LB}
O. L. Berman, Yu. E. Lozovik, D. W. Snoke, and R. D. Coalson, Phys. Rev. B
{\bf 70}, 235310 (2004); {\it ibid.} {\bf 73}, 235352 (2006); Solid State
Commun. {\bf 134}, 47 (2005); Physica E {\bf 34}, 268 (2006); J. Phys.:
Condens. Matter {\bf 19}, 386219 (2007).

\bibitem{rand}
Yu. E. Lozovik and A. M. Ruvinskii, JETP {\bf 87}, 788 (1998); Yu. E. Lozovik,
O. L. Berman, and A. M. Ruvinsky, JETP Lett. {\bf 69}, 616 (1999); Yu. E.
Lozovik and M. Willander, Appl. Phys. A {\bf 71}, 379 (2000).

\bibitem{screening}
V. M. Kovalev and A. V. Chaplik, JETP Lett. {\bf 92}, 185 (2010); M. Alloing,
A. Lema\^itre, and F. Dubin, Europhys. Lett. {\bf 93}, 17007 (2011).

\bibitem{prb046010193}
V. Srinivas, J. Hryniewicz, Y. J. Chen, and C. E. C. Wood, Phys. Rev. B {\bf
46}, 10193 (1992).

\bibitem{prb078045313}
C. Schindler and R. Zimmermann, Phys. Rev. B {\bf 78}, 045313 (2008).

\bibitem{prl099176403}
M. Combescot, O. Betbeder-Matibet, and R. Combescot, Phys. Rev. Lett. {\bf
99}, 176403 (2007).

\bibitem{pssc00302457}
A. Filinov, M. Bonitz, P. Ludwig, and Yu. E. Lozovik, Phys. Status Solidi C
{\bf 3}, 2457 (2006).

\bibitem{Tsuppr}
A. A. Dremin, V. B. Timofeev, A. V. Larionov, J. Hvam, and C. Soerensen, JETP
Lett. {\bf 76}, 450 (2002).

\bibitem{prl110216407}
R. Maezono, P. L\'opez R\'ios, T. Ogawa, R. J. Needs, Phys. Rev. Lett. {\bf 110},
216407 (2013).

\bibitem{jetpl0660355}
Yu. E. Lozovik, O. L. Berman, and V. G. Tsvetus, JETP Lett. {\bf 66}, 355
(1997).

\bibitem{jetp10600296}
Yu. E. Lozovik, I. L. Kurbakov, G. E. Astrakharchik, and M. Willander, JETP
{\bf 106}, 296 (2008).

\bibitem{trions}
M. V. Kochiev, V. A. Tsvetkov, and N. N. Sibeldin, JETP Lett. {\bf 95}, 481
(2012); M. D. Fraser, H. H. Tan, and C. Jagadish, Phys. Rev. B {\bf 84},
245318 (2011).

\bibitem{Tbec}
A. V. Gorbunov and V. B. Timofeev, JETP Lett. {\bf 84}, 329 (2006); V. B.
Timofeev and A. V. Gorbunov, J. Appl. Phys. {\bf 101}, 081708 (2007); Phys.
Status Solidi C {\bf 5}, 2379 (2008); J. Phys.: Conf. Ser. {\bf 148}, 012049
(2009).

\bibitem{prb075113301}
P. Pieri, D. Neilson, and G. C. Strinati, Phys. Rev. B {\bf 75}, 113301
(2007).

\bibitem{prl096227402}
A. T. Hammack, M. Griswold, L. V. Butov, L. E. Smallwood, A. L. Ivanov, and A.
C. Gossard, Phys. Rev. Lett. {\bf 96}, 227402 (2006).

\bibitem{prb083165308}
G. J. Schinner, E. Schubert, M. P. Stallhofer, J. P. Kotthaus, D. Schuh, A. K.
Rai, D. Reuter, A. D. Wieck, A. O. Govorov, Phys. Rev. B {\bf 83}, 165308
(2011).

\bibitem{jetpl0960138}
A. V. Gorbunov and V. B. Timofeev, JETP Lett. {\bf 96}, 138 (2012).

\bibitem{prl103087403}
A. A. High, A. K. Thomas, G. Grosso, M. Remeika, A. T. Hammack, A. D.
Meyertholen, M. M. Fogler, L. V. Butov, M. Hanson, and A. C. Gossard, Phys.
Rev. Lett. {\bf 103}, 087403 (2009).

\bibitem{prb056005306}
W. Zhao, P. Stenius, and A. Imamoglu, Phys. Rev. B {\bf 56}, 5306 (1997); M.
H. Szymanska, J. Keeling, and P. B. Littlewood, Phys. Rev. Lett. {\bf 96},
230602 (2006).

\bibitem{prl097187402}
S. Yang, A. T. Hammack, M. M. Fogler, L. V. Butov, and A. C. Gossard, Phys.
Rev. Lett. {\bf 97}, 187402 (2006).

\bibitem{exexp}
Y. Shilo, K. Cohen, B. Laikhtman, K. West, L. Pfeiffer, and R. Rapaport, Nat.
Comm. {\bf 4}, 2335 (2013); M. Stern, V. Umansky, and I. Bar-Joseph, Science
{\bf 343}, 55 (2014); M. Alloing, M. Beian, D. Fuster, Y. Gonzalez, L.
Gonzalez, R. Combescot, M. Combescot, and F. Dubin, Europhys. Lett. {\bf 107},
10012 (2014); S. Yang, L. V. Butov, B. D. Simons, K. L. Campman, and A. C.
Gossard, Phys. Rev. B {\bf 91}, 245302 (2015).

\bibitem{extheor}
K. I. Golden, G. J. Kalman, P. Hartmann, and Z. Donko, Phys. Rev. E {\bf 82},
036402 (2010); Y. G. Rubo and A. V. Kavokin, Phys. Rev. B {\bf 84}, 045309
(2011); A. V. Kavokin, M. Vladimirova, B. Jouault, T. C. H. Liew, J. R.
Leonard, and L. V. Butov, Phys. Rev. B {\bf 88}, 195309 (2013); D. Neilson, A.
Perali, and A. R. Hamilton, Phys. Rev. B {\bf 89}, 060502 (2014); S. V.
Andreev, A. A. Varlamov, and A. V. Kavokin, Phys. Rev. Lett. {\bf 112}, 036401
(2014); M. Combescot, R. Combescot, M. Alloing, and F. Dubin, Phys. Rev. Lett.
{\bf 114}, 090401 (2015); F.-C Wu, F. Xue, and A. H. MacDonald, Phys. Rev. B
{\bf 92}, 165121 (2015).

\bibitem{jpc011l00483}
A. V. Klyuchnik and Yu. E. Lozovik, J. Phys. C {\bf 11}, L483 (1978).

\bibitem{pla228000399}
Yu. E. Lozovik and A. V. Poushnov, Phys. Lett. A {\bf 228}, 399 (1997).

\bibitem{prb090165430}
A. K. Fedorov, I. L. Kurbakov, Yu. E. Lozovik, Phys. Rev. B {\bf 90}, 165430
(2014).

\bibitem{Andreev}
S. V. Andreev, Phys. Rev. Lett. {\bf 110}, 146401 (2013); Phys. Rev. B {\bf
92}, 041117 (2015).

\bibitem{jetpl0790473}
Yu. E. Lozovik, S. Yu. Volkov, and M. Willander, JETP Lett. {\bf 79}, 473
(2004).

\bibitem{SSma}
I. L. Kurbakov, Yu. E. Lozovik, G. E. Astrakharchik, and J. Boronat, Phys.
Rev. B {\bf 82}, 014508 (2010); J. Ye, J. Low Temp. Phys. {\bf 158}, 882
(2010); M. Matuszewski, T. Taylor, and A. V. Kavokin, Phys. Rev. Lett. {\bf
108}, 060401 (2012).

\bibitem{prl115075303}
A thermodynamically stable supersolid state can be realized when roton-like attraction \cite{prb090165430} and
many-body repulsion\cite{manybody} coexist, see Z.-K. Lu, Y. Li, D. S. Petrov, and G.
V. Shlyapnikov, Phys. Rev. Lett. {\bf 115}, 075303 (2015).

\bibitem{prl105070401}
A. Filinov, N. V. Prokof'ev, and M. Bonitz, Phys. Rev. Lett. {\bf 105}, 070401
(2010).

\bibitem{pla366000487}
Yu. E. Lozovik, I. L. Kurbakov, and M. Willander, Phys. Lett. A {\bf 366}, 487
(2007).

\bibitem{prl100250401}
E. B. Sonin, Phys. Rev. Lett. {\bf 102}, 106407 (2009); S. I. Shevchenko, Phys. Rev. B {\bf 56}, 10355 (1997).

\bibitem{QClum}
J. Keeling, L. S. Levitov, and P. B. Littlewood, Phys. Rev. Lett. {\bf 92},
176402 (2004); J. Ye, T. Shi, and L. Jiang, {\it ibid.} {\bf 103}, 177401
(2009).

\bibitem{ssc126000269}
Yu. E. Lozovik, I. L. Kurbakov, and I. V. Ovchinnikov, Solid State Commun.
{\bf 126}, 269 (2003).

\bibitem{prb091161413}
C.-E. Bardyn, T. Karzig, G. Refael, and T. C. H. Liew, Phys. Rev. B {\bf 91},
161413 (2015).

\bibitem{sr0005011925}
Q.-D. Jiang, Z.-Q. Bao, Q.-F. Sun, and X. C. Xie, Sci. Rep. {\bf 5}, 11925
(2015).

\bibitem{apl104162105}
T. Hakio\u{g}lu, E. \"Ozg\"un, and M. G\"unay, Appl. Phys. Lett. {\bf 104},
162105 (2014).

\bibitem{prl118127402}
R. Anankine, M. Beian, S. Dang, M. Alloing, E. Cambril, K. Merghem, C. G. Carbonell, A. Lemaitre, and F. Dubin, Phys. Rev. Lett. {\bf 118}, 127402 (2017).

\bibitem{EE}
J. P. Eisenstein and A. H. MacDonald, Nature {\bf 432}, 691 (2004).

\bibitem{graphen}
A. Perali, D. Neilson, and A. R. Hamilton, Phys. Rev. Lett. {\bf 110}, 146803
(2013); D. S. L. Abergel, M. Rodriguez-Vega, E. Rossi, and S. Das Sarma, Phys.
Rev. B {\bf 88}, 235402 (2013).

\bibitem{bandgap}
O. L. Berman, R. Ya. Kezerashvili, and K. Ziegler, Phys. Rev. B {\bf 85},
035418 (2012).

\bibitem{topol}
D. K. Efimkin, Yu. E. Lozovik, and A. A. Sokolik, Phys. Rev. B {\bf 86},
115436 (2012).

\bibitem{sr0005010354}
L. V. Kulik, A. V. Gorbunov, A. S. Zhuravlev, V. B. Timofeev, S. Dickmann, and
I. V Kukushkin, Sci. Rep. {\bf 5}, 10354 (2015).

\bibitem{14041418}
M. M. Fogler, L. V. Butov, and K. S. Novoselov, Nat. Comm. {\bf 5}, 4555
(2014); E. V. Calman, C. J. Dorow, M. M. Fogler, L. V. Butov, S. Hu, A.
Mishchenko, and A. K. Geim, Appl. Phys. Lett. {\bf 108}, 101901 (2016).


\bibitem{TMD}
K. F. Mak, C. Lee, J. Hone, J. Shan, and T. F. Heinz, Phys. Rev. Lett. {\bf
105}, 136805 (2010); A. K. Geim and I. V. Grigorieva, Nature {\bf 499}, 419
(2013).

\bibitem{nc0006006242} P. Rivera, J. R. Schaibley, A. M. Jones, J. S. Ross, S. F. Wu, G. Aivazian, P.
Klement, K. Seyler, G. Clark, N. J. Ghimire, J. Q. Yan, D. G. Mandrus, W. Yao,
and X. D. Xu, Nat. Comm. {\bf 6}, 6242 (2015).
\bibitem{sci351000688} P. Rivera, K. L. Seyler, H. Y. Yu, J. R. Schaibley, J. Q. Yan, D. G. Mandrus, W. Yao, and X. D. Xu, Science {\bf 351}, 688 (2016).

\bibitem{MoS2MoS2}
T. Cheiwchanchamnangij and W. R. L. Lambrecht, Phys. Rev. B {\bf 85}, 205302
(2012); S. F. Wu, J. S. Ross, G.-B. Liu, G. Aivazian, A. Jones, Z. Y. Fei, W.
G. Zhu, D. Xiao, W. Yao, D. Cobden, and X. D. Xu, Nat. Phys. {\bf 9}, 149
(2013).

\bibitem{jap099066104}
A. T. Hammack, N. A. Gippius, S. Yang, G. O. Andreev, L. V. Butov, M. Hanson,
and A. C. Gossard, J. Appl. Phys. {\bf 99}, 066104 (2006).

\bibitem{apl097201106}
Y. Y. Kuznetsova, A. A. High, and L. V. Butov, Appl. Phys. Lett. {\bf 97},
201106 (2010).

\bibitem{other}
Z. V\"or\"os, D. W. Snoke, L. Pfeiffer, and K. West, Phys. Rev. Lett. {\bf
97}, 016803 (2006); K. Kowalik-Seidl, X. P. V\"ogele, F. Seilmeier, D. Schuh,
W. Wegscheider, A. W. Holleitner, and J. P. Kotthaus, Phys. Rev. B {\bf 83},
081307 (2011); M. Alloing, A. Lema\^itre, E. Galopin, and F. Dubin, Sci. Rep.
{\bf 3}, 1578 (2013).

\bibitem{flat}
G. Chen, R. Rapaport, L. N. Pffeifer, K. West, P. M. Platzman, S. Simon, Z.
V\"or\"os, and D. Snoke, Phys. Rev. B {\bf 74}, 045309 (2006); K.
Kowalik-Seidl, X. P. V\"ogele, B. N. Rimpfl, G. J. Schinner, D. Schuh, W.
Wegscheider, A. W. Holleitner, and J. P. Kotthaus, Nano Lett. {\bf 12}, 326
(2012).

\bibitem{prl103016403}
Z. V\"or\"os, D. W. Snoke, L. Pfeiffer, and K. West, Phys. Rev. Lett. {\bf
103}, 016403 (2009).

\bibitem{elst}
M. Remeika, J. C. Graves, A. T. Hammack, A. D. Meyertholen, M. M. Fogler, L.
V. Butov, M. Hanson, and A. C. Gossard, Phys. Rev. Lett. {\bf 102}, 186803
(2009); M. Remeika, J. R. Leonard, C. J. Dorow, M. M. Fogler, L. V. Butov, M.
Hanson, and A. C. Gossard, Phys. Rev. B {\bf 92}, 115311 (2015).

\bibitem{apl100061103}
M. Remeika, M. M. Fogler, L. V. Butov, M. Hanson, and A. C. Gossard, Appl.
Phys. Lett. {\bf 100}, 061103 (2012).

\bibitem{magn}
A. Abdelrahman and B. S. Ham, Phys. Rev. B {\bf 86}, 085445 (2012); {\it
ibid.} {\bf 87}, 125311 (2013).


\bibitem{prl106196806}
A. G. Winbow, J. R. Leonard, M. Remeika, Y. Y. Kuznetsova, A. A. High, A. T.
Hammack, L. V. Butov, J. Wilkes, A. A. Guenther, A. L. Ivanov, M. Hanson, and
A. C. Gossard, Phys. Rev. Lett. {\bf 106}, 196806 (2011).

\bibitem{ol0032002466}
A. A. High, A. T. Hammack, L. V. Butov, M. Hanson, and A. C. Gossard, Opt.
Lett. {\bf 32}, 2466 (2007).

\bibitem{apl085005830}
J. Krauss, J. P. Kotthaus, A. Wixforth, M. Hanson, D. C. Driscoll, A. C.
Gossard, D. Schuh, M. Bichler, Appl. Phys. Lett. {\bf 85}, 5830 (2004).

\bibitem{jap117023108}
M. W. Hasling, Y. Y. Kuznetsova, P. Andreakou, J. R. Leonard, E. V. Calman,
C. J. Dorow, L. V. Butov, M. Hanson, and A. C. Gossard, J. Appl. Phys. {\bf
117}, 023108 (2015).

\bibitem{ol0040000589}
Y. Y. Kuznetsova, P. Andreakou, M. W. Hasling, J. R. Leonard, E. V. Calman, L.
V. Butov, M. Hanson, and A. C. Gossard, Opt. Lett. {\bf 40}, 589 (2015).

\bibitem{grad}
J. R. Leonard, M. Remeika, M. K. Chu, Y. Y. Kuznetsova, A. A. High, L. V.
Butov, J. Wilkes, M. Hanson, and A. C. Gossard, Appl. Phys. Lett. {\bf 100},
231106 (2012); P. Andreakou, S. V. Poltavtsev, J. R. Leonard, E. V. Calman, M.
Remeika, Y. Y. Kuznetsova, L. V. Butov, J. Wilkes, M. Hanson, and A. C.
Gossard, {\it ibid.} {\bf 104}, 091101 (2014); C. J. Dorow, Y. Y. Kuznetsova,
J. R. Leonard, M. K. Chu, L. V. Butov, J. Wilkes, M. Hanson, and A. C.
Gossard, {\it ibid.} {\bf 108}, 073502 (2016).

\bibitem{prl099047602}
J. Rudolph, R. Hey, and P. V. Santos, Phys. Rev. Lett. {\bf 99}, 047602 (2007);
S. Lazi\'c, A. Violante, K. Cohen, R. Hey, R. Rapaport, and P. V. Santos,
Phys. Rev. B {\bf 89}, 085313 (2014).

\bibitem{prl105116402}
E. A. Cerda-M\'endez, D. N. Krizhanovskii, M. Wouters, R. Bradley, K.
Biermann, K. Guda, R. Hey, P. V. Santos, D. Sarkar, and M. S. Skolnick, Phys.
Rev. Lett. {\bf 105}, 116402 (2010).

\bibitem{pra084053601}
S. M\"{u}ller, J. Billy, E. A. L. Henn, H. Kadau, A. Griesmaier, M.
Jona-Lasinio, L. Santos and T. Pfau, Phys. Rev. A {\bf 84}, 053601 (2011).

\bibitem{pra079043623}
N. Fabbri, D. Cl$\grave{e}$ment, L. Fallani, C. Fort, M. Modugno, K. M. R. van
der Stam, and M. Inguscio, Phys. Rev. A {\bf 79}, 043623 (2009).

\bibitem{prl114055301}
L.-C. Ha, L. W. Clark, C. V. Parker, B. M. Anderson, and C. Chin, Phys. Rev.
Lett. {\bf 114}, 055301 (2015).

\bibitem{np0006000056}
P. T. Ernst, S. G\"{o}tze, J. S. Krauser, K. Pyka, D.-S. L\"{u}hmann, D.
Pfannkuche and K. Sengstock, Nat. Phys. {\bf 6}, 56 (2009).

%\bibitem{njp012083025}
%X. Du, S. Wan, E. Yesilada, C. Ryu, D. J. Heinzen, Z. Liang and B. Wu, New J.
%Phys., {\bf 12}, 083025 (2010).

\bibitem{epjd027000247}
M. Kr\"amer, C. Menotti, L. Pitaevskii, and S. Stringari, Eur. Phys. J. D {\bf
27}, 247 (2003).

\bibitem{pra058001480}
K. Berg-S$\o$rensen and K. M$\o$lmer, Phys. Rev. A {\bf 58}, 1480(1998).

\bibitem{SSBH}
K. G\'oral, L. Santos, and M. Lewenstein, Phys. Rev. Lett. {\bf 88}, 170406
(2002); H. P. B\"uchler and G. Blatter, {\it ibid.} {\bf 91}, 130404 (2003);
C. Trefzger, C. Menotti, and M. Lewenstein, {\it ibid.} {\bf 103}, 035304
(2009); I. Danshita and Carlos A. R. S\'a de Melo, {\it ibid.} {\bf 103},
225301 (2009).

\bibitem{pra083023623}
R. M. Wilson and J. L. Bohn, Phys. Rev. A {\bf 83}, 023623 (2011).

\bibitem{meff}
J. Javanainen, Phys. Rev. A {\bf 60}, 4902 (1999); M. Kr\"amer, L. Pitaevskii,
and S. Stringari, Phys. Rev. Lett. {\bf 88}, 180404 (2002).

\bibitem{prl069000644}
K. Huang and H.-F. Meng, Phys. Rev. Lett. {\bf 69}, 644 (1992).

\bibitem{pit-str}
S. Giorgini, L. Pitaevskii, and S. Stringari, Phys. Rev. B {\bf 49}, 12938
(1994).

\bibitem{pr0104000576}
O. Penrose and L. Onsager, Physical Review {\bf 104}, 576 (1956).

\bibitem{rmp047000331} A. J. Leggett, Rev. Mod. Phys. {\bf 47}, 331 (1975).
\bibitem{volovik} G.E. Volovik, Superfluid He3. Hydrodynamics and inhomogeneous states, Sov. Sci. Rev. Sect. A: Physics reviews, {\bf 1}, 23-84 (1979).
    Ed. Khalatnikov I.M., London, UK: Harwood Academic Publishers, 1979, xv+305 pp. ISBN: 3-7186-0004-8;
\bibitem{nphys008000253} V. P. Mineev, Nat. Phys. {\bf 8}, 253 (2012).

\bibitem{pra086043612}
J.-S. You, H. Lee, S. Fang, M. A. Cazalilla, and D.-W. Wang, Phys. Rev. A {\bf
86}, 043612 (2012).

%\bibitem{prl081003108} D. Jaksch, C. Bruder, J.I. Cirac, C.W. Gardiner, P. Zoller, Phys. Rev. Lett. {\bf 81}, 3108 (1998).
%\bibitem{nature415000039} M. Greiner, O. Mandel, T. Esslinger, T.W. H\"ansch,I. Bloch, Nature {\bf 415}, 39 (2002).

\bibitem{prl103165301}
M. Iskin and C. A. R. S\'a de Melo, Phys. Rev. Lett. {\bf 103}, 165301 (2009).

\bibitem{prb084205113}
D. Chowdhury, E. Berg, and S. Sachdev, Phys. Rev. B {\bf 84}, 205113 (2011).

\bibitem{pra084033625}
A. Macia, F. Mazzanti, J. Boronat, and R. E. Zillich, Phys. Rev. A {\bf 84},
033625 (2011).

\bibitem{pra086053602}
C. Ticknor, Phys. Rev. A {\bf 86}, 053602 (2012).

\bibitem{pra089023605}
J. Sch\"onmeier-Kromer and L. Pollet, Phys. Rev. A {\bf 89}, 023605 (2014).

\bibitem{prb044004503}
P. Minnhagen and P. Olsson, Phys. Rev. B {\bf 44}, 4503 (1991).

\bibitem{pla376000480}
P. Muruganandam and S. K. Adhikari, Phys. Lett. A {\bf 376}, 480 (2012); G.
Bismut, B. Laburthe-Tolra, E. Marechal, P. Pedri, O. Gorceix, and L. Vernac,
Phys. Rev. Lett., ${\bf 109}$, 15, 155302 (2012).

\bibitem{prl106065301}
C. Ticknor, R. M. Wilson, and J. L. Bohn, Phys. Rev. Lett. {\bf 106}, 065301
(2011).

\bibitem{njp005000050}
D. M. Stamper-Kurn, New J. Phys. {\bf 5}, 50 (2003).

\bibitem{prl111170402}
B. C. Mulkerin, R. M. W. van Bijnen, D. H. J. O'Dell, A. M. Martin, and N. G.
Parker, Phys. Rev. Lett. {\bf 111}, 170402 (2013).

\bibitem{sr0006019380}
X.-F. Zhang, L. Wen, C.-Q. Dai, R.-F. Dong, H.-F. Jiang, H. Chang, and S.-G.
Zhang, Sci. Rep. {\bf 6}, 19380 (2016).

\bibitem{pra008001111} M. E. Fisher, M. N. Barber, D. Jasnow, Phys. Rev. A {\bf 8}, 1111 (1973).

\bibitem{weakcorr}Strong correlation mode ($[N-N_0]/N \sim 1$) is frequently realized for two-dimensional excitons \cite{ssc144000399}. However, we present only solution for weak correlations because it allows one to obtain general formulas in an analytical form. We assume that qualitative conclusions we obtain can be applied to strongly correlated systems.
\bibitem{ignoringspin}
For strictly bosonic excitons spin degrees of freedom are manifest only in the splitting (Zeeman, exchange, etc.) of the excitonic band bottom. One can show by a straightforward calculation using the Gross-Pitaevskii equation that in the weak correlation and modulation regime condensate populates only the lowest in energy spin degree of freedom. The condensate spin basis coincides then with the single-exciton spin basis allowing one to consider the problem assuming a single spin band.

\bibitem{loz-yud} Yu.E. Lozovik, V.I. Yudson, Physica A {\bf 93}, 493 (1978).
\bibitem{nc_finite}The finiteness of (\ref{nc_dens}) means that (\ref{nc_dens}) will be sufficiently small for weak potentials $U$ and $V_0$, which means that $n_0=n-n'$ will certainly be positive.
\bibitem{alpha_dep} Notice that expressions for $n-n_0$ and $\langle \hat H ' \rangle$ are invariant under ${\bf q}\leftrightarrow-{\bf q}$. For quantities carrying $+$/$-$ indices (except $f_{\pm}$) this operation as it can be seen from (\ref{bogolsol}) is equivalent to simply $+ \leftrightarrow -$. In the expressions for observables there are only two vector quantities --- ${\bf P}$ and ${\bf q}$. ${\bf P}$ enters all the expressions except for the spectrum (\ref{spectrum}) in the form of $\alpha$. However, because in (\ref{spectrum}) ${\bf P}$ forms a product with the integration variable (for energy) the answer will contain product with some other vector. All the other ${\bf P}$ are part of  $\alpha$ so it means that in the final answer ${\bf P}$ will come in $\alpha$. ${\bf q}$ in the answer can be in the form of ${\bf q}^2$ or $\alpha$. Let us decompose the answer in $\alpha$ (remember that ${\bf P}$ is infinitesimal). Coefficients going with $\alpha$ in odd powers should be zero because the answer is symmetric under ${\bf q} \leftrightarrow -{\bf q}$. In the vicinity of ${\bf P}=0$ this leads to (\ref{nc_decomp}).


\bibitem{anis-estim}
To make quantitative estimates we have used (\ref{Ysmatr}) and (\ref{Deltas}), but not (\ref{csnd}). This allows one to avoid ambiguity between (\ref{y_sq}) and (\ref{csnd}), which differ due to a neglected $O(\Delta_s^2)$ term in the first expression.


\bibitem{prb042009225}
A. Alexandrou, J. A. Kash, E. E. Mendez, M. Zachau, J. M. Hong, T. Fukuzawa,
and Y. Hase, Phys. Rev. B {\bf 42} 9225 (1990).
\bibitem{prb072075428} R. Rapaport, G. Chen, S. Simon, O. Mitrofanov, L. Pfeiffer, and P. M. Platzman, Phys. Rev. B ${\bf 72}$, 075428 (2005).

\bibitem{pr0155000080}
J. W. Kane and L. P. Kadanoff, Phys. Rev. {\bf 155}, 80 (1967).

\bibitem{Popov}
V. N. Popov, {\it Functional Integrals in Quantum Field Theory and Statistical
Physics} (D. Reidel, Dordrecht, 1983).

\bibitem{qft-hd}
H.-F. Meng, Phys. Rev. B {\bf 49}, 1205 (1994). A brief overview of the hydrodynamic field theory calculation of the density matrix is given in \cite{jetp10600296} for uniform 2D systems at $T=0$.  A more detailed account is presented for the 1D case in D. L. Luxat and A. Griffin, Phys. Rev. A {\bf 67}, 043603 (2003).

\bibitem{angleresolvedlumin}
Expression (\ref{Iangle}) is valid for the case of a direct optical transition and applying rescaling \cite{pra086043612,prb044004503} $x\to x/\eta$, $y\to\eta y$, where $\eta=[({\rm Y}_s)_{xx}/({\rm Y}_s)_{yy}]^{1/4}$, to the long-wavelength density matrix asymptotics \cite{pr0155000080} $\rho_1(r)\propto r^{-\alpha}$.

\bibitem{T>0}
In 2D quasicondensate at $T=0$ becomes a ususal BEC\cite{pr0155000080}. Anisotropy of the angle-resolved luminescence of a BEC is determined by the geometrical form of the order parameter and is not connected to the superfluidity anisotropy.

\bibitem{prb062001548} L. V. Butov, A. V. Mintsev, Yu. E. Lozovik, K. L. Campman, and A. C. Gossard, Phys. Rev. B ${\bf 62}$, 1548 (2000).
\bibitem{prl087216804} L.V. Butov, C.W. Lai, D. S. Chemla, Yu. E. Lozovik, K. L. Campman, and A. C. Gossard, Phys. Rev. Lett. {\bf 87}, 216804 (2001).
\bibitem{ssc144000399} Yu. E. Lozovik, I.L. Kurbakov, G.E. Astrakharchik, J. Boronat, M. Willander, Solid State Comm. {\bf 144}, 399 (2007).
\bibitem{prl079000553}
M. R. Andrews, D. M. Kurn, H.-J. Miesner, D. S. Durfee, C. G. Townsend, S.
Inouye, and W. Ketterle, Phys. Rev. Lett. {\bf 79}, 553 (1997); {\bf 80}, 2967
(1998).
\bibitem{prl094226401}
Z. V\"or\"os, R. Balili, D. W. Snoke, L. Pfeiffer, and K. West, Phys. Rev.
Lett. {\bf 94}, 226401 (2005).



\bibitem{collrel}
C. Ciuti, V. Savona, C. Piermarocchi, A. Quattropani, and P. Schwendimann, Phys. Rev. B {\bf 58}, 7926 (1998).

\bibitem{phonrel}
J. Lee, E. S. Koteles, and M. O. Vassell, Phys. Rev. B {\bf 33}, 5512 (1986);
R. K. Basu and P. Ray, {\it ibid.} {\bf 45}, 1907 (1992).

\bibitem{nonzero}
Strictly speaking, in the regime of spatially resolved cw pump a finite exciton recombination rate already yields a non-zero motion velocity in the form of a continouous inflow to the studied area of the QW. We neglect this effect in (\ref{v0=0}) assuming the lifetime to be large enough.


\bibitem{10ps}$\tau_{\rm dissip}^{\rm norm}$ can be estimated as the inverse damping rate of elementary excitations due to disorder. This rate has been found in \cite{LB} to be greater than $0.1$meV/$\hbar$ thus $\tau_{\rm dissip}^{\rm norm}\lesssim10$ ps.

\bibitem{jap104063515} A. G. Winbow, L. V. Butov, and A. C. Gossard, J. Appl. Phys. ${\bf 104}$, 063515 (2008).

\bibitem{prb050001119} H. Shi, G. Verechaka and A. Griffin, Phys. Rev. B ${\bf 50}$, 1119 (1994).
\bibitem{prl084000806}K. W. Madison, F. Chevy, W. Wohlleben, and J. Dalibard, Phys. Rev. Lett. {\bf 84}, 806 (2000)).





%\bibitem{prb045011403} R. Eccleston, B. F. Feuerbacher, J. Kuhl, W. W. R\"uhle, and K. Ploog, Phys.Rev. B, ${\bf 45}$, 19, 11403 (1992).



%\bibitem{prl088130402} S. Giovanazzi, D. O$'$Dell, and G. Kurizki, Phys. Rev. Lett., {\bf 88}, 130402 (2002).
%\bibitem{prb082014508} I. L. Kurbakov, Yu. E. Lozovik, G. E. Astrakharchik, and J. Boronat, Phys. Rev. B, {\bf 82}, 014508 (2010).
%\bibitem{prl090250403} L. Santos, G.V. Shlyapnikov, and M. Lewenstein, Phys. Rev. Lett., {\bf 90}, 250403 (2003).
%J. Ye, Journal of Low Temperature Physics, {\bf 158}, 5, 882 (2010).
%R. Mottl, F. Brennecke, K. Baumann, R. Landig, T. Donner, T. Esslinger, Science, {\bf 336}, 6088, 1570 (2012);
%M. Matuszewski, T. Taylor and A. V. Kavokin, Phys. Rev. Lett., {\bf 108}, 060401 (2012).

\bibitem{jetpl0940800} A. V. Gorbunov, V. B. Timofeev, and D. A. Demin, JETP Lett. {\bf 94}, 800 (2012).

\bibitem{prl098060405}
G. E. Astrakharchik, J. Boronat, I. L. Kurbakov, and Yu. E. Lozovik, Phys.Rev. Lett. {\bf 98}, 060405 (2007).

\bibitem{ns}
The BKT transition temperature in an anisotropic superfluid has the form\cite{BKT} $T_c\approx\pi\hbar^2n_s/2m$, where\cite{pra086043612} $n_s/m=\sqrt{({\rm Y}_s)_{xx}({\rm Y}_s)_{yy}}$,  and $({\rm Y}_s)_{xx}$, $({\rm Y}_s)_{yy}$ are the diagonal elements of the helicity modulus tensor. Taking $n\approx n_0$ into account one obtains from (\ref{Ysmatr}) $n_s=n_0\sqrt{1-2\Delta_s}$.

%\bibitem{pra061063608}
%In the Bogoliubov limit the density-density correlator tends to its hydrodynamical value even at scales less than the healing length,
%F. Zambelli, L. Pitaevskii, D. M. Stamper-Kurn, and S. Stringari, Phys. Rev. A
%{\bf 61}, 063608 (2000).
%\bibitem{prb052001236}
%J. Boronat, J. Casulleras, F. Dalfovo, S. Stringari, and S. Moroni, Phys. Rev.
%B {\bf 52}, 1236 (1995).


%\bibitem{prl098060405}
%G. E. Astrakharchik, J. Boronat, I. L. Kurbakov, and Yu. E. Lozovik, Phys. Rev. Lett. {\bf 98}, 060405 (2007).

\bibitem{narrowchannel}
When an uncompressible non-dissipatively moving liquid flows into a narrow channel, Bernoulli's law can lead to arbitrary high velocities even if outside the channel flow velocity is small. Thus the flow velocity in the channel can reach Landau critical velocity or higher values. The flow through the channel can be created by, e.g., pumping the excitons only on one side of the channel, while on the other they will only recombine.


\bibitem{notunnel}
Because $E_z^{\rm av}$ is sufficiently weak (see Tab.~\ref{t:efield}), we
neglect tunneling electron (hole) out of the QW.

\bibitem{prb088075434}
S. Sim, J. Park, J.-G. Song, C. In, Y.-S. Lee, H. Kim, and H. Choi, Phys. Rev.
B {\bf 88}, 075434 (2013).

\bibitem{tau}
For the MoS$_2$/hBN structure, where we assume that the in-plane magnetic
field is absent (see Sec.~\ref{estimation}). Therefore, the bottom of
excitonic dispersion is inside the radiative zone such that
$\tau\sim\tau_{\rm bright}$. However, for GaAs-based structures the necessary
in-plane magnetic field $B_{\parallel}=8$ T (see Tab. \ref{t:move}) is so
large that the bottom of excitonic dispersion is far out of the radiative
zone. Thus for moderate densities and sufficiently low temperatures the {\it
ab initio} modeling \cite{ssc144000399} yields an occupation of radiative zone
so low, that the main recombination channel is nonradiative (with momentum
transfer to an additional particle). Exciton lifetime with respect to this
channel is proportional \cite{prb040001074} to $M^2$ and \cite{prb050001119}
$\tau_{\rm dir}$, i.e., $\tau\propto\tau_{\rm bright}$ (see (\ref{taubright}).
A lower bound for the nonradiative exciton recombination time in the limit of
strong in-plane magnetic fields is \cite{prb062015323}
$\tau\sim50\;\tau_{\rm bright}$.

\bibitem{FTT} Yu.E. Lozovik, V.N. Nishanov, Fiz. Tverd. Tela {\bf 18}, 3267 (1976).

\bibitem{uncertrel}
Here we employ uncertainty relation $\Delta p\sim\hbar/\Delta x$. The ground
state energy of an exciton in an $e$-$h$ bilayer is given by (\ref{HDx}).
Minimizing (\ref{HDx}) with respect to $\Delta x$, we obtain the estimate
$r_{\rm ex}=\Delta x_m$ for exciton radius and $E_B=-H(\Delta x_m)$ for
exciton binding energy.

\bibitem{pra081013612}
G. E. Astrakharchik, J. Boronat, I. L. Kurbakov, Yu. E. Lozovik, and F.
Mazzanti, Phys. Rev. A {\bf 81}, 013612 (2010).

\bibitem{sci335000947}
Unlike the MoS$_2$/hBN/MoS$_2$ structure, MoSe$_2$/hBN/WSe$_2$ is a type II QW.  In this case the results of the tunneling model used in \ref{1DSE&EB} do not agree well with the experimental data\cite{sci351000688}. To estimate $\tau$ for MoSe$_2$/hBN/WSe$_2$ CQWs we use instead a crude estimate based on the experiment L. Britnell, R. V. Gorbachev, R. Jalil, B. D. Belle, F. Schedin, A. Mishchenko, T. Georgiou, M. I. Katsnelson, L. Eaves,
S. V. Morozov, N. M. R. Peres, J. Leist, A. K. Geim, K. S. Novoselov, and L. A. Ponomarenko, Science {\bf 335}, 947 (2012). For a non-resonant tunneling process this estimate yields $\tau$ 6 orders of magnitude larger then for a zero thickness hBN barrier (2 orders of magnitude per hBN monolayer). Thus at least $\tau>0.1$ $\mu$s.

\bibitem{srep05722}
 S. Ge, X. Liu, X. Qiao, Q. Wang, Z. Xu, J. Qiu, P. Tan, J. Zhao and D. Sun, Sci. Rep. {\bf 4}, 5722 (2014).

\bibitem{prb090045422} Zhenghe Jin, Xiaodong Li, Jeffrey T. Mullen, and Ki Wook Kim Phys. Rev. B {\bf 90}, 045422
(2014).

\bibitem{manybody}
B. Laikhtman and R. Rapaport, Phys. Rev. B {\bf 80}, 195313 (2009); G. J.
Schinner, J. Repp, E. Schubert, A. K. Rai, D. Reuter, A. D. Wieck, A. O.
Govorov, A. W. Holleitner, and J. P. Kotthaus, Phys. Rev. B {\bf 87}, 205302
(2013).

\bibitem{prb040001074}
R. Ferreira and G. Bastard, Phys. Rev. B {\bf 40}, 1074 (1989).

\bibitem{prb062015323}
A. Parlangeli, P. C. M. Christianen, J. C. Maan, I. V. Tokatly, C. B.
Soerensen, and P. E. Lindelof, Phys. Rev. B {\bf 62}, 15323 (2000).



\end{thebibliography}
\end{document}